\documentclass[prb, twocolumn, superscriptaddress, notitlepage, nofootinbib, flotfix, longbibliography]{revtex4-1}
\usepackage{amsmath,amsfonts, amssymb, amsthm, dsfont}
\usepackage{yfonts}
\usepackage{bm}
\usepackage{mathrsfs}
\usepackage{graphicx}
\usepackage{verbatim}
\usepackage{hyperref}
\usepackage{wasysym}
\usepackage{xcolor}
\usepackage{tikz}
\usepackage{bbold}
\usepackage[caption=false]{subfig}

\newcommand{\ket}[1]{|#1\rangle}

\renewcommand{\ol}[1]{\overline{#1}}
\newcommand{\comments}[1]{}
\newcommand{\mb}[1]{\mathbf{#1}}
\renewcommand{\cal}[1]{\mathcal{#1}}
\renewcommand{\Ref}[1]{[\onlinecite{#1}]}

\newcommand{\spd}[1]{(#1+1)d}

\def\U{\mathrm{U}(1)}
\def\H{\mathcal{H}}
\def\Z{\mathbb{Z}}
\def\TT{\mathsf{T}}

\def\id{\mathds{1}}

\makeatletter
\def\l@subsubsection#1#2{}
\makeatother

\begin{document}
\title{Gapped boundary of \spd{4} beyond-cohomology bosonic SPT phase}
\author{Xinping Yang}
\author{Meng Cheng}
\affiliation{Department of Physics, Yale University, New Haven, CT 06520-8120, USA}

\begin{abstract}
    In this work we study gapped boundary states of $\Z_N$ bosonic symmetry-protected topological (SPT) phases in \spd{4}, which are characterized by mixed $\Z_N$-gravity response, and the closely related phases protected by $C_N$ rotation symmetry. We show that if $N\notin \{2,4,8,16\}$, any symmetry-preserving boundary theory is necessarily gapless for the root SPT state. We then propose a \spd{3} $\Z_2$ gauge theory coupled to fermionic matter as a candidate boundary theory for $N=2,4,8,16$, where the anomalous symmetry is implemented by invertible topological defects obtained from gauging \spd{2} chiral topological superconductors. For the $C_N$ case, we present an explicit construction for the boundary states for $N=2,4,8,16$, and argue that the construction fails for other values of $N$. 
\end{abstract}
\maketitle

\section{Introduction}
The concept of bulk-boundary correspondence is fundamental to the theory of topological phases. It is most well-understood when the bulk is an invertible symmetry-protected topological (SPT) phase, where the boundary has an 't Hooft anomaly of the symmetry group that protects the bulk SPT phase~\cite{Chen_2013}. The presence of a 't Hooft anomaly puts nontrivial constraints on the low-energy dynamics, and in particular excludes a trivially gapped symmetric ground state. More generally, any low-energy theory realized in the system must have the given anomaly. Generally, three options are possible for the boundary theory: gapless, symmetry breaking or a symmetry-preserving topologically ordered phase (when the boundary has spatial dimension $D$ greater than 1). 

The last possibility, namely the boundary forming a symmetry-enriched topological (SET) phase~\cite{VishwanathSenthil, XieAnomalousSurface} with the 't Hooft anomaly, has been extensively investigated in the past few years. In particular for $D=2$, general theories of SET phases in both bosonic and fermionic systems have been formulated~\cite{SET, TeoSET2015, TarantinoSET2016, BulmashPRB2022a, BulmashPRB2022b, Aasen:2021vva}. Systematic methods to compute 't Hooft anomalies given a SET phase have been given. It is also known that certain 't Hooft anomalies can not be matched by any SET in $D=2$, thus any symmetry-preserving theory must be gapless. Known examples of ``symmetry-enforced gaplessness" involve continuous and anti-unitary symmetry group~\footnote{There are also examples involving finite higher-form symmetries, see e.g. \Ref{CordovaOhmori1}.}, such as a bosonic anomaly for $\mathrm{SO}(5)\times\Z_2^\TT$ symmetry~\cite{WangDQCP2017,GeraedtsPRB2017}.

The focus of this work is $D=3$, where a full theory of SET phases is not available yet. A necessary ingredient of such a theory is a complete understanding of the structure of 't Hooft anomaly. It is well-known that 't Hooft anomalies are classified by SPT phases in one dimension higher. Interestingly, in \spd{4} there is a class of bosonic SPT phases protected by unitary symmetry, that goes beyond the well-known ``group-cohomology" classification. Such ``beyond-cohomology" SPT phases can be understood as decorating lower-dimensional invertible topological phases on symmetry defects. The physical characterization of these beyond-cohomology SPT phases turns out to be rather subtle.  An argument based on topological quantum field theory (TQFT) consideration suggests that for $N=2$, the nontrivial phase can be characterized by the ground state having an odd $\Z_2$ charge when put on the $\mathbb{CP}^2$ manifold. However, a commuting-projector model Hamiltonian, unitarily equivalent to a group-cohomology SPT model in flat space, can also exhibit the same phenomenon~\cite{Fidkowski_2020}. Therefore a definite invariant of the phase requires considering the boundary anomaly (or the closely related defect decoration)~\cite{Fidkowski4DSPT}.

We will study the boundary theory of the root $\Z_N$ beyond-cohomology SPT phase, using two complementary points of view. First of all, we use a theorem proven by Cordova and Ohmori to show that only when $N=2,4,8,16$ , there can be $Z_N$ symmetry-preserving TQFT boundary states. We then propose a boundary TQFT for such allowed values of $N$: a \spd{3} $\Z_2$ gauge theory with a fermionic $\Z_2$ charge (which will be referred to as a fermionic $\Z_2$ gauge theory from now on). Secondly, we provide an explicit construction of the symmetry-preserving boundary state for $N=2,4,8,16$. However, given that solvable models for $\Z_N$ BC SPT phases are still lacking for $N>2$, we turn to a different but related system, that is a \spd{4} SPT phase protected by $C_N$ rotation symmetry. Following the dimensional reduction approach~\cite{SongPRX2017}, we show that our construction of gapped boundary topological orders for $N=2,4,8$ preserves the boundary $C_N$ symmetry, while surpringly fails for $N=16$. Yet a slight modification yields a similar boundary state.

\section{Bosonic $\Z_N$ SPT in \spd{4}}
First we review the classification of \spd{4} bosonic SPT phases, following ~\cite{Xiong:2016deb, Gaiotto:2017zba}.
Let $G$ be a compact unitary group. The ``group-cohomology" SPT phases are classified by  $\H^5[G, \U]$, and the ``beyond-cohomology" SPT phases are classified by $\H^2[G, \Z]$, as we will argue below. The total group of SPT phases is an extension of $\H^2[G, \Z]$ by $\H^5[G, \U]$. We show that for $3\nmid N$, the group structure is $\Z_N^2$, while for $3\mid N$, it is $\Z_{3N}\times \Z_{N/3}$. Details to determine the group structure of $\Z_N$ SPT are written in Appendix \ref{group-structure}. For finite $G$, exactly solvable models (either in the form of a state-sum TQFT, or commuting-projector Hamiltonian) are known for such phases~\cite{Chen_2013}. 

\subsection{Defect decoration}
\label{domain wall}
To see how to construct the $\Z_N$ BC SPT phase, it would be instructive to start from $G=\U$ and then break it down to $\Z_N$. The group-cohomology SPT phases are classified by $\H^5[\U, \U]=\Z$, which can be characterized by the \spd{4} quantum Hall response. The beyond-cohomology phases ($\H^2[\U, \Z]=\H^1[\U,\U]=\Z$) can be constructed as follows: suppose the $\U$ symmetry is spontaneously broken so the system is in a superfluid phase. In a \spd{4} superfluid, the vortices are codimension-2 domain walls (i.e. spatially they are surfaces). It is well-known that by proliferating the vortices one can restore the $\U$ symmetry and enter an insulating state. To create a nontrivial BC SPT state, the vortices are decorated by \spd{2} invertible states. Once the U(1) symmetry is broken down to $\Z_N$, a U(1) vortex should be viewed as a junction fusing $N$ fundamental $\Z_N$ domain walls together.
   
   The ``beyond-cohomology" SPT phases can thus be understood as decorating \spd{2} nontrivial invertible states on junctions of symmetry defects~\cite{2014NatCo...5.3507C, Wang:2021nrp}. Recall that invertible phases in \spd{2} form a $\Z$ group, generated by the so-called $E_8$ state with chiral central charge $c_-=8$. It has the simplest edge theory being a chiral $(E_8)_1$ conformal field theory (CFT). We then consider decorating the codimension-2 tri-junctions of symmetry defects, which are surfaces in 4D space, with invertible states. A tri-junction is labeled by a pair of group elements $(\mb{g,h})\in G$. Fusing this pair of defects gives a $\mb{gh}$ domain wall in $G$. The decoration pattern is parametrized by an integer-valued function $n(\mb{g,h})\in \Z$, evaluated at each tri-junction. Physically $n(\mb{g,h}) = c_-/8$ is the $E_8$ state decorated on the tri-junctions. In order to get a short-range entangled state, it is necessary that the decoration patterns on defect configurations which can be locally deformed to each other are (adiabatically) equivalent. This requirement is equivalent to imposing the 2-cocycle condition on $n(\mb{g}, \mb{h})$ when we consider a junction that fuses three symmetry defects $\mb{g,h,k}$ into $\mb{ghk}$. In addition, the following procedure does not change the underlying phase of matter: we create a pair of invertible states labeled by $m(\mb{g})$ and $-m(\mb{g})$ on the $\mb{g}$ defect and move them to the adjacent tri-junctions, i.e. $n(\mb{g,h})\rightarrow n(\mb{g,h})+m(\mb{g})+m(\mb{h})-m(\mb{gh})$. Thus $n(\mb{g},\mb{h})$ is a 2-cocycle defined up to 1-coboundary and its classification is given by $\H^2[G, \Z]$.

Mathemtically, we can understand such defect decoration by looking at the explicit expression of the 2-cocycle. For $G=\Z_N$, a 2-cocycle in $\H^2[\Z_N, \Z]$ can be written in this canonical form 
\begin{equation}
  n(a,b)=\frac{s}{N}(a+b-[a+b]_N),  
  \label{nab}
\end{equation}
where $a,b\in \{0,1,\dots, N-1\}$ denote the elements of $\Z_N$ additively, $s$ takes values in $\{0,1,\dots, N-1\}$, and $[x]_N$ means $x$ mod $N$. This explicit expression of $n$ shows that fusing $N$ of the fundamental $\Z_N$ defects should yield a \spd{2} invertible state labeled by $n(1,1)+n(2,1)+\cdots + n(N-1,1)=s$.

A key question here is what \spd{2} invertible states can be decorated consistently on the $\U$ vortex sheets or the junction of $\Z_N$ defects. Naively one might think of the $E_8$ states but it is not obvious that such a decoration is consistent. While we do not have a direct way to check the consistency at the level of a wavefunction, it is useful to consider the following QFT arugment: assuming that the system can be described by a relativistic field theory, then we can study the theory on a general curved manifold (in Euclidean spacetime) and its response to background $\U$ gauge field. That is, we consider the partition function $\cal{Z}(M_5, A)$ of the theory defined on a closed 5-manifold $M_5$ equipped with a $\U$ background gauge field $A$. To write down the well-defined action, it is convenient to introduce a 6-dimensional manifold $B_6$ with $\partial B_6=M_5$. The gauge field $A$ is also extended to $B_6$. The BC SPT phase is characterized by the following topological term~\cite{Wang:2014pma, Lapa:2016kkv, WenPRB2015}:
\begin{equation}
    \cal{Z}(B_6, A) = \exp \left(ik\int_{B_6} {F}\wedge p_1\right).
    \label{ZU1}
\end{equation}
where $F=dA$ is the field strength, and $p_1$ is the Pontryagin class of the tangent bundle of the manifold. Then the partition function on $M_5$ can be written as $\cal{Z}(M_5,A)= \text{exp}(2 \pi i k\int_{M_5}A\wedge p_1)$. Notice that a well-defined topological term in Eq.\ref{ZU1} should be independent of the choice of manifold extension. In other words, the right-hand side of Eq. \eqref{ZU1}, when evaluated on any closed 6-manifold, must give $1$. This requires $k$ to be an integer~\footnote{This is because $\int\frac{{F}}{2\pi}\wedge p_1$ is always an integer}. On the other hand, observe that $e^{2\pi i k \int p_1}$ defines an invertible theory in \spd{3}, \footnote{For a closed 4-manifold, $\int_{M_4}{p_1}=3{\sigma}(M_4)$ where $\sigma(M_4)$ is the signature of the 4-manifold. The partition function $e^{2\pi ik\int p_1}=e^{6\pi ik\sigma}=e^{\frac{2\pi i}{8}\cdot 24k\cdot \sigma}$, } thus when the manifold has a \spd{2} boundary, the theory reduces to a boundary gravitational Chern-Simons term with chiral central charge $24k$. Thus we identify $24k$ as the chiral central charge of the invertible theory decorated on the domain wall junctions~\cite{WenPRB2015}. In other words, the index $s$ in Eq. \eqref{nab} must be a multiple of 3.

This QFT argument suggests that one can only decorate minimally $c_-=24$ invertible states (i.e. three copies of $E_8$ states) on vortex surfaces. However, it is worth emphasizing that we have assumed relativistic symmetry in this argument, so it is not entirely clear whether the same argument applies to gapped phases in non-relativistic systems, such as lattice models. A related fact is that the ground state wavefunctions of such an invertible state with $c_-$ a multiple of 24 on any closed surfaces are completely invariant under modular transformation (no additional phase factor). Such modular invariance may be required for a consistent decoration. 

\subsection{Partition function for $\Z_N$ BC SPT Phase}
The argument in the previous section does not fully capture the subtlety of writing down the response action of $\Z_N$ BC SPT phase. When $3\nmid N$, the response action is indeed
\begin{equation}
    \cal{Z}(M_5, A)=\exp\left(\frac{2\pi ik}{N}\int_{M_5}A\cup p_1\right), k\in \Z_N.
    \label{Z_ZN}
\end{equation}
where $A$ is the background $\Z_N$ gauge field (valued in $\Z/N\Z$). 

The $3\mid N$ case requires a separate treatment.  In the case of $N=3$ it is known that the action \eqref{Z_ZN} turns out to be equivalent to that of a $\Z_3$ group-cohomology SPT phase~\footnote{We thank C.-M. Jian for crucial discussions on this point.}. In fact, the $\Z_3$ SPT phases are classified by $\Z_9$~\cite{Wan:2018bns}, where the generator is the root BC SPT phase. We discuss the partition function of the root phase in Appendix. \ref{Z3}.

In the Hamiltonian formalism, one can interpret this action as follows: let $M_5=M_4\times S^1$, where $M_4$ is a closed 4-manifold. We also assume there is  a unit $\Z_N$ holonomy along $S^1$. For $3\nmid N$, the partition function on $M_5$ evaluates to $e^{\frac{6\pi ik}{N}\sigma(M_4)}$. The physical interpretation is that the ground state on a closed 4-manifold $M_4$ has $\Z_N$ charge $3k\sigma(M_4)$ mod $N$. 
However the ground state charge becomes ambiguious away from the pure TQFT limit. The sublety was recently examplified in the generalized double semion model~\cite{Fidkowski_2020}, which is an exactly solvable (commuting-projector) lattice model with the same ground state property as the $\Z_2$ BC SPT phase on any closed 4-manifold, but on the other hand is locally equivalent to a group-cohomology $\Z_2$ SPT model.

\subsection{Constraints on the boundary states}
We are interested in the existence of gapped, symmetry-preserving \spd{3} boundary conditions for the BC $\Z_N$ SPT phases. 
Recently, within the mathematical framework of TQFTs, [\onlinecite{CordovaOhmori1, CordovaOhmori2}] established a necessary condition for a \spd{3} 't Hooft anomaly to be saturated by a symmetry-preserving TQFT: the corresponding \spd{4} SPT topological partition function must evaluate to 1 on $K_3\times S^1$ (with any choice of background gauge field)~\footnote{Cordova-Ohmori's theorem is proved for spin (i.e. fermionic) TQFTs. However, upon a close examination of the proof, one finds that the proof only requires a spin manifold, such as $K_3$, in order to apply the Wall's theorem, but the argument does not require the theory being fermionic/spin. Thus the theorem applies to bosonic systems as well.}. Here $K_3$ is a closed simply-connected 4-manifold with signature $16$. In other words, if one can find a gauge field configuration of the corresponding symmetry such that the partition function yields a phase factor different from 1 on $K_3\times S^1$, then the SPT phase can not have a symmetry-preserving TQFT boundary.  

Let us use the criterion to study the $\Z_N$ BC SPT phase, whose partition function is given by Eq. \eqref{Z_ZN}. If we require the partition function to be $1$ on $K_3\times S^1$, for $3\nmid N$ using \eqref{Z_ZN} we find
\begin{equation}
 \frac{48k}{N}\in \Z.   
\end{equation}
 For $k=1$, it means a symmetry-preserving TQFT boundary is possible only for
\begin{equation}
    N=2,4,8,16.
\end{equation}

The case of $3\mid N$ is more delicate. We provide an argument that the partition function on $K_3\times S^1$ is not 1 for the root $N=3$ BC SPT phase in the Appendix \ref{Z3}, and hence there can not be a symmetry-preserving boundary TQFT. Interestingly, even though the root phase does not allow symmetry-preserving gapped boundary, 3 copies of the root phase is equivalent to a group-cohomology SPT phase, which can have symmetric gapped boundary. Similar results can be proven for other $3\mid N$.

\section{Bosonic $C_N$ SPT in \spd{4}}
\label{CN SPT}
A drawback of our discussions of the $\Z_N$ BC SPT phases is that it is entirely based on topological partition functions, and at the moment we do not have a concrete microscopic model for them. In this section we turn to a different but closely related symmetry $C_N$, the point group of $N$-fold rotations, and study SPT phases protected by this symmetry. The advantage of considering spatial SPT states (i.e. those protected by spatial symmetries) is that they can be classified and explicitly constructed using the block construction~\cite{SongPRX2017}. It is valid to relate $\Z_N$ SPT phases with $C_N$ SPT phases because there is a one-to-one correspondence between these two SPT phases, known as the crystalline correspondence principle~\cite{ThorngrenElse2018}. 

Let us elaborate on the relation between the $\Z_N$ SPT phases and the $C_N$ ones. Starting from a $\Z_N$ SPT phase, we create a symmetry-breaking state in the following way: insert $N$ copies of $\Z_N$ domain walls in a $C_N$ symmetric configuration. 
This system breaks $\Z_N$ and $C_N$, but preserves the diagonal subgroup called $C_N'$. At the rotation center, the $N$ domain walls fuse together to a codimension-2 defect, which is the state that lives at the rotation center. Therefore, from a $\Z_N$ SPT state we can always construct a $C_N$ SPT state. For the other direction, consider a continuum QFT with continuous spatial symmetry (i.e. SO($D$) and translations, where $D$ is the spatial dimension). The $C_N$ rotation can always be written as a $\Z_N$ internal symmetry transformation combined with the corresponding rotation in SO$(D)$. It is then expected that the theory with the $\Z_N$ symmetry is in the corresponding $\Z_N$ SPT phase. 

\subsection{Block construction for $C_N$ SPT}
  
  Let us carry out the block construction for $C_N$ symmetry in \spd{4}. By the definition of SPT phases,  the bulk state can be disentangled everywhere except on the rotation ``axis", which is two-dimensional in 4D, and the $C_N$ symmetry reduces to a $\Z_N$ internal symmetry on the rotation axis. Now there are two possibilities: a $\Z_N$ group-cohomology SPT phase, or a $E_8$ state, on which the $\Z_N$ does not act. The latter corresponds to the generator of the $\Z_N$ beyond-cohomology SPT phases.  In Appendix \ref{group-structure} we compute the group structure of the $C_N$ SPT phases, which turns out to be $\Z_N\times \Z_N$ for $3\nmid N$, and $\Z_{3N}\times\Z_{N/3}$ for $3\mid N$. Also note that in this construction for $C_N$ SPT, the rotation center can be decorated with any \spd{2} invertible topological phase. Unlike the internal $\Z_N$ case, there is no need to impose the $c_-=24k$ condition from the boundary gravitational Chern-Simons terms of $\Z_N$ topological action.

  \begin{figure}
    \centering
    \includegraphics[width=\columnwidth]{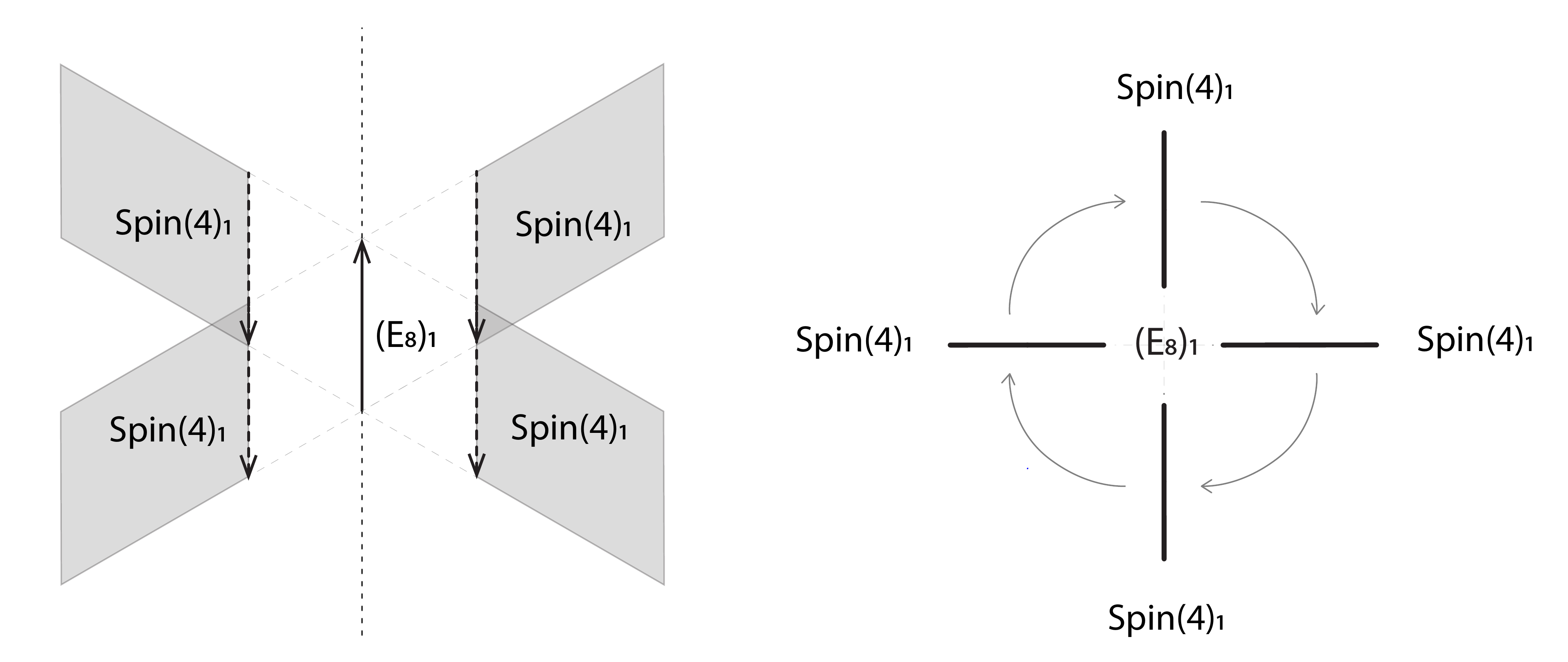}
    \caption{Illustration of the construction for a gapped state on the boundary of \spd{4} $C_N$ BC SPT state.}
    \label{fig:Z4}
\end{figure}

We then choose a 3D boundary perpendicular to the rotation axis, so the boundary is invariant under rotation. The rotation plane in the bulk terminates as the 1D axis on the boundary. Since there is an $E_8$ state on the plane, the 1D axis carries the corresponding $(E_8)_1$ chiral edge mode. In the following we fix that the rotation axis to be in the $z$ direction. 

In order to create a fully gapped \spd{3} boundary, we use the following construction: choose $N$ half planes all terminating at the $z$ axis, the positions of which are related to each other by $C_N$ rotation. For example, one of them could be the plane defined by $y=0, x\geq 0$, and the others are obtained by $C_N$ rotations. On each plane we place a 2D chiral topological phase $\cal{B}$. Again all of them are placed in a $C_N$-symmetric way. At the 1D rotation center, we have $N$ edge modes from the topological phases on the half-planes and the $(E_8)_1$ CFT from the rotation center in the \spd{4} bulk. The setup is illustrated in Fig. \ref{fig:Z4} for $N=4$. We require that these edge modes together can be gapped out while preserving the $C_N$ symmetry. In other words, the $N$ blocks have a $C_N$-preserving gapped boundary to $E_8$. We can further fold the $N$ layers into one topological phase denoted by $\mathcal{B}^{\boxtimes N}$, where the $C_N$ symmetry becomes the $\Z_N$ cyclic permutation symmetry between layers. A similar method was used in Ref. [\onlinecite{Qi:2017qui}] to study \spd{2} topological phases enriched by reflection symmetry, which is essentially the $N=2$ case.

To summarize, our construction requires topological phases that satisfy the following conditions:
\begin{enumerate}
\item The topological phase $\mathcal{B}$ has a chiral central charge $c_-=\frac{8}{N}$.
    \item  $N$ layers of $\mathcal{B}$ can have a fully gapped edge to a $E_8$ state.
    \item The gapped edge preserves the $\Z_N$ cyclic permutation symmetry of the $N$ layers.
\end{enumerate}
We will say $\cal{B}$ is \emph{$N$-gappable} if all the conditions are satisfied.

Therefore, the construction reduces to finding a $\Z_N$ symmetry-preserving gapped boundary (to an $E_8$ state) of $N$ copies of $\cal{B}$. 

\subsection{Gapped boundary conditions}
\label{constraint}
Below we will study this problem using the mathematical framework of modular tensor category (MTC), also known as the anyon theory in physics literature~\cite{Kitaev:2005hzj}. In this formalism, a topologically ordered phase in \spd{2} is fully described in terms of the universal data of the low-energy quasi-particle excitations, i.e. the anyons. The universal data describe the fusion and braiding properties of the anyons. Alternatively, this collection of data also suffices to specify the \spd{2} TQFT associated with the topological phase. We should note that the MTC description does not fully determine the edge property, i.e. the chiral central charge $c_-$. In fact, it can be shown that the MTC (or the anyon theory) determines $c_-$ mod 8. Physically the ambiguity precisely comes from stacking $E_8$ states in the bulk, which does not affect the anyon excitations but can change edge $c_-$ by integer multiples of 8. 

Gapped boundaries of a topological phase can also be described in this formalism~\cite{kitaev2012, eliens2013, levin2013, kong2014, lan2015, NeupertPRB2016, Cong2017}. Each gapped boundary is associated to a unique (composite) anyon object, which determines which anyons can condense on the boundary. This object is called the Lagrangian algebra, denoted by $\mathcal{A}$ below.  
For Abelian anyons, the condensed anyons form a Lagrangian subgroup~\cite{levin2013}. In this case, it is relatively simple to state the condition for a set of anyons to condense: they must all be bosons, and the mutual braiding statistics between them must all be trivial. In addition, the number of anyons in the Lagrangian group must be the square root of the total number of anyons. The definition of the algebra in the general case is reviewed in Appendix \ref{app:condensation}.

We can now define the notion of Witt equivalence for topological phases. Two topological phases $\cal{B}_1$ and $\cal{B}_2$ are Witt equivalent, if $\cal{B}_1\boxtimes \ol{\cal{B}}_2$ has a fully gapped interface to an invertible state (i.e. some copies of $E_8$ states). Here 
 $\boxtimes$ denotes the operation of stacking two systems, and $\ol{\cal{B}}_2$ is the mirror image of $\cal{B}_2$. In other words, there is a gapped interface between $\cal{B}_1$ and $\cal{B}_2$ as long as we are allowed to freely stack copies of $E_8$ states. Mathematically, two MTCs $\cal{B}_1$ and $\cal{B}_2$ are Witt equivalent if $\cal{B}_1\boxtimes \ol{\cal{B}}_2$ has a Lagrangian algebra. The Witt equivalence classes of MTCs form an Abelian group, known as the Witt group of MTCs. 

With these definitions, the first two conditions of $N$-gappability implies that the MTC $\cal{B}$ has a minimal order $N$ in the Witt group. However, it is also known that the structure of the Witt group is highly constrained~\cite{DGNO, DGNOWitt1, DGNOWitt2}: the order of elements in the Witt group can not be any odd integer. In fact, the only possible finite values are $2^n$ with $1\leq n\leq 5$ (there are obviously elements of infinite order). From this point of view, we can immediately rule out any odd $N\geq 3$ in the construction.

In our problem there is also a global symmetry $G$, i.e. the $\Z_N$ group of cyclic permutations of the $N$ layers, and the gapped boundary must preserve this global symmetry. Therefore it is necessary to incorporate global symmetry in the categorical formalism. First of all, the global symmetry can act on the anyons in nontrivial ways. In this case, the action is given by the layer permutation. There are more subtle aspects of symmetry actions in the topological phase but they do not occur in our system, so we will not go into details. For more details on symmetry-enriched topological phases in \spd{2}, see Ref. [\onlinecite{SET}].

Next, we will need to understand whether a given gapped boundary, or the Lagrangian algebra, can preserve the global symmetry. Clearly, the set of condensed anyons must be invariant under the global symmetry (layer permutations in our case), otherwise the symmetry is explicitly broken. For a condensed anyon $a$, we define $G_a$ as the subgroup of $G$ that keeps $a$ invariant. Then the condensed anyon $a$ should carry a well-defined charge under the symmetry group $G_a$. Here by ``charge" we mean a one-dimensional representation of $G_a$, i.e. a homomorphism from $G_a$ to U(1). Different choices of these charges correspond to different types of symmetry-preserving gapped boundary conditions. Again we note that here our description is heuristic, and a more precise formulation is given in Appendix \ref{app:condensation}.

To understand the physical consequence of the symmetry charges of the condensed anyons, it is useful to consider the nature of the gapped boundary. By definition, the gapped boundary is the interface between the topological phase and an invertible state, which is often taken to be the vacuum. However, in the presence of global symmetry, the invertible state could be a nontrivial SPT state. The nature of this invertible state is determined by the symmetry charges of the condensate~\cite{ChengPRR2020, Qi:2017qui}. Here we are mainly interested in the case where the invertible state has no nontrivial SPT order. Following Ref. [\onlinecite{Qi:2017qui}], we study the problem by gauging the $\Z_N$ symmetry in the $\cal{B}^{\boxtimes N}$ theory, the result of which is denoted by $[\cal{B}^{\boxtimes N}]/\Z_N$. The interface then becomes one between the gauged theory $[\cal{B}^{\boxtimes N}]/\Z_N$, and a $\Z_N$ gauge theory obtained from gauging the $\Z_N$ SPT state (possibly stacked with an $E_8$ state). After gauging, the Lagrangian algebra in $\cal{B}^{\boxtimes N}$ is ``lifted" to one in $[\cal{B}^{\otimes N}]/\Z_N$, however the lifting requires additional data, i.e. the symmetry charges of the condensed anyons. Once the Lagrangian algebra and its lifting are given, one can apply the theory of anyon condensation to determine the nature of the $\Z_N$ gauge theory after condensation. We leave the details of the derivations in Appendix \ref{app:gauging}.

Below we apply this theory to several examples. In particular, we will consider a family of anyon theories known as Kitaev's 16-fold ways~\cite{Kitaev:2005hzj}. They can be described as \spd{2} fermionic topological superconductors with Chern number $\nu$ coupled to $\Z_2$ gauge field, or as Spin$(\nu)_1$ Chern-Simons theory. We will show that Spin$(\nu)_1$ is $\frac{16}{(16,\nu)}$-gappable when $\nu$ is even. However, we find that Spin$(\nu)_1$ is \emph{not} 16-gappable when $\nu$ is odd. Instead, we find a different but closely related construction for the $C_{16}$ case.

\subsubsection{Spin$(2n)_1$}
In this section we will consider Spin$(2n)_1$ theories.  First we review the basic properties of these MTCs.
Spin$(2n)_1$ has four anyons: $1, \psi, v, v'=v\times\psi$. $\psi$ is a fermion and satisfies $\psi^2=1$. $v$ can be viewed as a fermion parity flux since the braiding phase between $v$ and $\psi$ is $M_{v\psi}=-1$. The topological twist factor of $v$ is $\theta_v=e^{\frac{i\pi n}{4}}$.  
The chiral central charge is $c_-=n$. The order of Spin$(2n)_1$ in the Witt group is $r=\frac{8}{\text{gcd}(8,n)}$. 

We will show that the Spin$(2n)_1$ MTCs are $r$-gappable.

First, we construct the Lagrangian subgroup for Spin$(2n)_1^{\boxtimes r}$, preserving the $\Z_r$ cyclic permutation symmetry.  We label the anyons by a $r$-tuple $(a_1,a_2,\cdots, a_r)$. Since all Spin$(2n)_1$ theories have total quantum dimension $\mathcal{D}=2$, the Lagrangian subgroup has to have size $2^r$.  Consider all bosons of the form
\begin{equation}
    \cal{A}_0=\{(a_1,a_2,\cdots, a_r)|a_i\in\{\id,\psi\}\},
\end{equation}
with an even number of $\psi$'s in the tuple. We will refer those as fermion bound states. There are $2^{r-1}$ such bosons. Then we fuse $(v,v,\cdots,v)$ with the fermion bound states, which is equivalent to replacing an even number of $v$'s with $v'=v\psi$. Together they generate a Lagrangian subgroup consist of $2^r$ bosons, and it is straightforward to check that they have trivial mutual braiding statistics, thus forming a Lagrangian subgroup.  Cleary this subgroup is invariant under the $\Z_r$ symmetry.  

As described in the beginning of this section, we need to examine finer structures of the Lagrangian algebra under the symmetry action. We perform these calculations carefully for two representative values of $n=1,2$ in Appendix \ref{app:condensation}. As discussed above, to determine the nature of the condensed phase it is necessary to know the symmetry charges of the condensed anyons. We find that a consistent choice is to have all the $\Z_r$-invariant anyons (i.e. $(\psi,\cdots,\psi)$ and $(v,\cdots,v)$) carry trivial charges under $\Z_r$. With this choice, we can apply the results in Appendix \ref{app:gauging} to show that the condensation leads to an $E_8$ state where the $\Z_r$ symmetry acts trivially. Together we have established that there is $\Z_r$-symmetric gapped interface between Spin$(2n)_1^{\boxtimes r}$ to an $E_8$ state where the $\Z_r$ symmetry acts trivially, so Spin$(2n)_1$ is $r$-gappable.

\subsubsection{Spin$(2n+1)_1$}
\label{Ising}

Let us now turn to Spin$(2n+1)_1$ theories.  Recall that the Spin$(2n+1)_1$ MTC has three types of anyons $\id, \sigma$ and $\psi$, where $\psi$ is a fermion and $\sigma$ is a non-Abelian anyon that satisfies $\sigma\times\sigma=\id+\psi$ and with a topological twist factor $\theta_\sigma=e^{\frac{i\pi(2n+1)}{8}}$.

We first enumerate all bosons in $[\text{Spin}(2n+1)_1]^{\boxtimes 16}$. There are $2^{15}$ Abelian bosons, which are bound states with an even number of fermions. There is also a non-Abelian boson $(\sigma,\cdots,\sigma)$, with quantum dimension $(\sqrt{2})^{16}=256$. We can form the following Lagrangian algebra:
\begin{equation}
    \mathcal{A}=\sum_{a \in \mathcal{A}_0}a + 128 (\sigma,\sigma,\cdots,\sigma).
\end{equation}
Unfortunately, due to the large multiplicity $128$, we are not able to obtain an explicit structure of the condensate. Therefore we adopt a different approach here. We condense only the Abelian subset $\mathcal{A}_0$ of $\mathcal{A}$ to obtain a $\Z_2$ toric code (TC) phase. From there, we try to construct a $C_{16}$ symmetry-preserving gapped boundary of the $\Z_2$ TC. We will show that in fact it is not possible to condense $\cal{A}$ in $[\text{Spin}(2n+1)_1]^{\boxtimes 16}$ without breaking the $\Z_{16}$ symmetry. This can be easily proven using the anyon condensation theory, see Appendix \ref{app:gauging}. Below we provide a more physical argument, which will also suggest a way to fix the problem.

We first condense the Abelian subgroup $\mathcal{A}_0$ of $\cal{A}$, resulting in a $\Z_2$ toric code (TC) phase. It is easy to see the only deconfined anyons are the Abelian anyons, and $(\sigma,\dots,\sigma)$. All the Abelian bosons are already condensed, and all the (Abelian) fermions are identified and become the same $\psi$ of the toric code phase. While $(\sigma,\cdots,\sigma)$ is deconfined, it is invariant under fusion with any of the condensed bosons, so it must split into direct sums of $e$ and $m$.

In the following denote $g$ as the cyclic permutation of the 16 layers. Namely, $g$ generates the $\Z_{16}$ symmetry group.  
 The $\Z_{16}$ symmetry should be preserved by the condensation, so the $\Z_2$ TC is enriched by the $\Z_{16}$ symmetry. Following the general classification~\cite{SET}, first we need to know how the generator $g$ permutes anyon types. Given that the permutation must preserve the fusion and braiding properties of anyons, there are two possibilities for the $\Z_2$ TC: either $g$ does not permute, or $g$ swaps $e$ and $m$. In the latter case, there is no symmetric gapped boundary. This is because the only Lagrangian subgroups for the $\Z_2$ TC are $\id+e$ and $\id+m$ and neither of them is invariant under $e\leftrightarrow m$. Therefore, a symmetric gapped boundary requires that $g$ does not permute $e$ and $m$ anyons.

To further determine the action of $g$ on anyon $\psi$, we take a slightly different approach. The $\Z_2$ toric code can be obtained by gauging the 2D fermion parity of 16 copies of $p_x+ip_y$ superconductors. In the block construction, we place the 16 copies in a $C_{16}$-invariant configuration, all terminating at the rotation axis. Then we mirror-fold them into a stack of $p_x+ip_y$ superconductors. The $g$ symmetry again permutes the 16 layers cyclically. In this construction, the $e$ and $m$ anyons correspond to a fermion parity flux.  The fermion parity flux binds 16 Majorana zero modes $\gamma_i$ for $i=1,2,\dots,16$, one from each layer. The local fermion parity of the flux is thus $P=\prod_i \gamma_i$. After gauging, the fermion parity flux with even and odd local fermion parity ($P=\pm 1$) become the $e$ and $m$ anyons.

Now under the $g$ action, $\gamma_i\rightarrow \gamma_{[i+1]_{16}}$, so we find 
\begin{equation}
    P=\prod_i\gamma_i\rightarrow \gamma_2\cdots\gamma_{16}\gamma_1=-P.
\end{equation}
Therefore, the $g$ symmetry flips the fermion parity of the flux, which becomes the $e\leftrightarrow m$ symmetry after gauging. As a result, there is no symmetric gapped boundary. 

However, we can fix this issue by modifying the transformation of the Majorana modes to the following form:
\begin{equation}
\begin{gathered}
    \gamma_i\rightarrow \gamma_{i+1}, i=1,\dots,15\\
    \gamma_{16}\rightarrow -\gamma_1.
\end{gathered}
\end{equation}
Under this transformation, the local parity $P$ remains unchanged. However, it then follows that $g^{16}$ acts as $\gamma_i\rightarrow -\gamma_i$ for all $i$, i.e. $g^{16}=(-1)^{N_f}$, so the fermion should transform projectively under $g$.

We thus conclude that the $\psi$ anyon should have $g^{16}=-1$ in order to avoid the $g$ symmetry swapping $e$ and $m$. Since $\psi=e\times m$, one of them must have $g^{16}=1$, which  can then condense to obtain a symmetric gapped boundary.  We have thus shown that there should exist a $C_{16}$-symmetric gapped boundary of the $\Z_2$ toric code phase to an invertible state.


Notice that in the original construction, we have 16 layers of Ising theories, and the $C_{16}$ rotation acts as the cyclic permutation. One can show that there is no nontrivial symmetry fractionalization for this symmetry.  It is natural to postulate that after condensing $\cal{A}_0$ one is led precisely to the toric code with $g^{16}=1$ on $\psi$. We can then conclude that the 16 layers of Ising topological orders admit a $C_{16}$-symmetric gapped boundary at the rotation center to an invertible state.

However we are not able to precisely determine the nature of the invertible state besides its chiral central charge $c_-=8$. So this \spd{3} fermionic $\Z_2$ gauge theory may exist on the boundary of the root $C_{16}$ BC SPT state, possibly stacked with another (in-cohomology) SPT state.

\section{\spd{3} Boundary TQFT}
 \label{symmetry in Z2 gauge theory}
 Ref. [\onlinecite{Fidkowski4DSPT}] constructed a fermionic $\Z_2$ gauge theory on the boundary of the $\Z_2$ BC SPT state. Here a fermionic $\Z_2$ gauge theory refers to a $\Z_2$ gauge theory with fermionic $\Z_2$ gauge charges. We will argue that the same theory can realize the anomaly for $\Z_N$ BC SPT state when  $N=4,8,16$ as well. Recall the domain wall decoration construction for $\Z_N$ BC SPT in \ref{domain wall}, the codimension-2 tri-junctions in the bulk are decorated by minimally three copies of $E_8$ states. Thus the truncated domain walls on the  \spd{3} boundary should host the $E_8$ on the junctions as well. 
 We will show that a fermionic $\Z_2$ gauge theory in \spd{3} has an anomalous $\Z_{16}$ (0-form) global symmetry~\cite{Johnson-Freyd:2020twl}. More precisely, the \spd{3} boundary TQFT is a fermionic $\Z_2$ gauge theory with $E_8$ defects.

 We now construct the codimension-1 invertible topological defects that implement the $\Z_{16}$ 0-form symmetry. We insert a \spd{2} chiral topological superconductor (TSC) of Chern number $\nu$ (equivalent to $\nu$ copies of $p+ip$ superconductors) into the fermionic theory before gauging, and then couple the system to a $\Z_2$ gauge field. This way we obtain an invertible topological defect of codimension-1 in the fermionic $\Z_2$ gauge theory~\footnote{One subtlety is that because of the chiral central charge, strictly speaking the defect is not completely topological, in the same sense as the \spd{2} Chern-Simons theory is not a topological field theory due to the framing anomaly. For example, the partition function with defect insertions depend on the geometry. However, since the dependence comes in as a phase factor, it does not affect the symmetry action.}, which defines a 0-form symmetry. From this construction naively it seems that the defect is labeled by the integer $\nu$ and fusion of two defects of $\nu_1$ and $\nu_2$ results in a defect of $\nu_1+\nu_2$. However, we will argue that $\nu$ is defined mod 16. This is because before gauging, a $\nu=16$ TSC is topologically equivalent to a $E_8$ state stacked with completely trivial gapped fermions. Since this equivalence can be generated by adiabatic evolution with a gapped local Hamiltonian preserving fermion parity, it is expected that the equivalence is preserved after gauging. In other words, a $\nu=16$ defect is equivalent to a $E_8$ state. Since the $E_8$ state is purely bosonic and decoupled from the $\Z_2$ gauge theory, when viewed as a topological defect it can only act on the $\Z_2$ gauge theory trivially. Therefore the faithful symmetry group generated by these TSC defects is $\Z_{16}$. However, the fact that 16 $\nu=1$ defects fuse to a $E_8$ state suggests that the $\Z_{16}$ symmetry is anomalous. Indeed this is exactly what should happen on the boundary of a $\Z_{16}$ BC SPT state~\footnote{Naively, 16 $\nu=1$ defects fuse to a $E_8$ state with $c_-=8$, but fusing 16 $\Z_{16}$ defects in the bulk leaves an invertible state with $c_-=24k$. However, it should be kept in mind that the chiral central charge of the defect in the \spd{3} theory is only defined mod 8, so more precisely 16 $\nu=1$ defects fuse to an invertible state with $c_-=8(16m+1)$, where $m$ is an integer. By choosing e.g. $m=2$, we have $k=11$ in the bulk.}.

  \begin{figure}
    \centering
    \includegraphics[width=0.9\columnwidth]{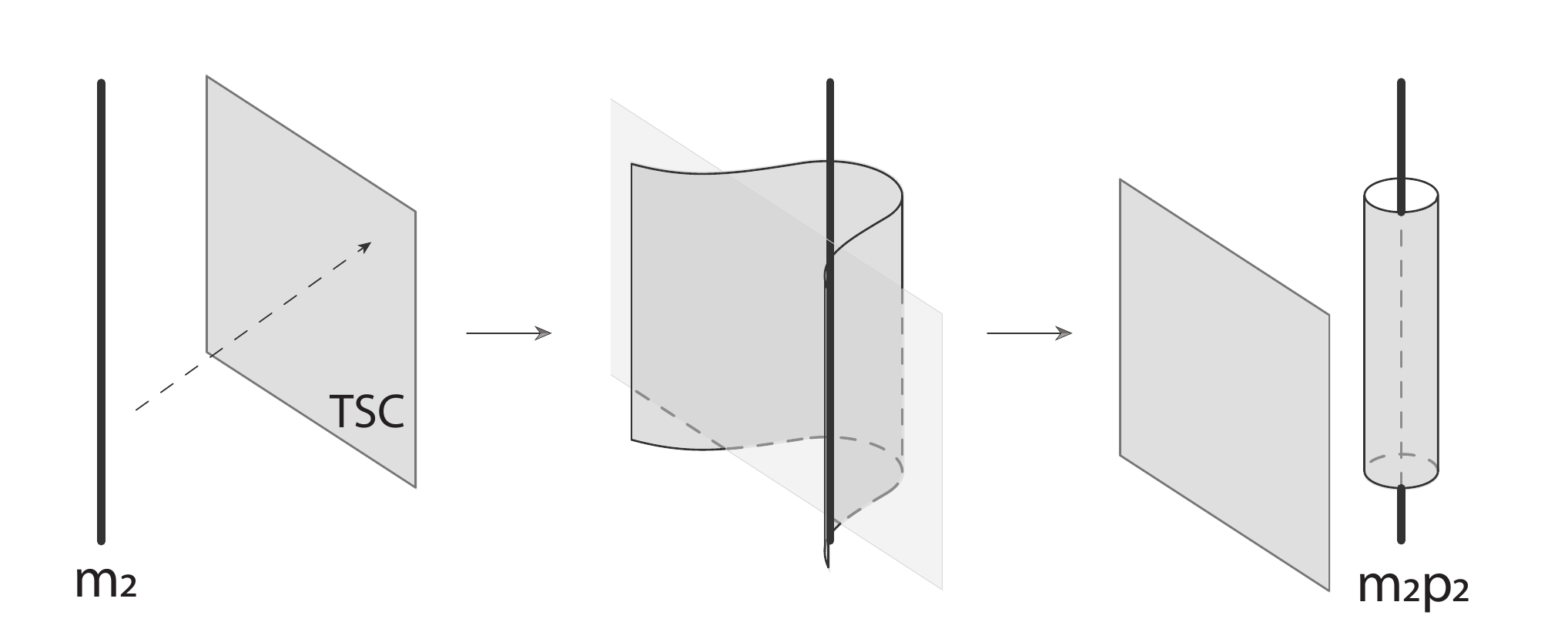}
    \caption{Illustration of the 0-form symmetry action on a flux loop.}
    \label{fig:loop}
\end{figure}

Let us examine how the symmetry acts on various objects in the theory. First we review the low-energy excitations of the $\Z_2$ gauge theory. There are two elementary types of excitations: a fermionic $\Z_2$ particle, and a $\Z_2$ flux loop (denoted by $m_2$ below, where the subscript 2 is the codimension). In addition, it is useful to introduce an invertible line defect as follow: We can think of the theory as a system of fermions in a gapped trivial state coupled to a $\Z_2$ gauge field. We insert a Majorana chain in the ungauged fermion system, and then gauge the $\Z_2$ fermion parity. The Majorana chain then becomes an invertible topological defect of codimension 2, which will be denoted by $p_2$. It is evident that the TSC defect does not act on the fermionic particle, and the nontrivial action only happens on the flux loops.
To see what is going on when a flux loop $m_2$ passes through the domain wall, we note that the process is equivalent to wrapping the topological superconductor around the flux loop. It is a well-known fact that when $\nu$ is odd, the topological superconductor when wrapped on a cylinder with anti-periodic boundary condition for the fermions (i.e. with a $\pi$ flux threading the cylinder) is equivalent to a Majorana chain~\cite{Read:1999fn}. See Fig. \ref{fig:loop} for an illustration of the symmetry action. Thus we find that $m_2\rightarrow m_2p_2$ when passing through a TSC defect with an odd $\nu$~\cite{Johnson-Freyd:2020twl}. 

 When $\nu$ is even, the symmetry does not change the type of the $m_2$ loop. Instead, let us consider a Hopf link of two flux loops, and pass the link through the domain wall. During this process, the worldlines of the four intersection points of the link with the domain wall precisely trace out the Hopf link, which correspond to a full braiding between two of them and result in a phase factor $\pm e^{\frac{i\pi\nu}{4}}$. Here the sign ambiguity $\pm$ comes from possible fermions attached to the flux loops. Notice that this characterization only applies to $\nu\equiv 2$ mod $4$. More generally, we can consider a ``three-loop braiding" process~\cite{WangPRL2014}, where two flux loops are linked to a base loop of the defect. The exchange statistics of the two flux loops is $e^{\frac{i\pi\nu}{8}}$, which can distinguish all different $\nu$ mod 16.

Now we describe the boundary TQFT for $C_N$ BC SPT states. Starting from the block construction of the boundary state in \ref{CN SPT}, we can now construct a $\Z_2$ gauge theory in the following way: fill the \spd{3} boundary with a fermionic $\Z_2$ gauge theory where the $C_r$ symmetry acts trivially (besides the coordinate transformation). The \spd{2} intrinsic topological orders (i.e. Spin$(2n)_1$ layers) need to be transformed into invertible codimension-1 defects in the $\Z_2$ gauge theory such that boundary theory is a \spd{3} TQFT. To achieve this, on each of the Spin$(2n)_1$ layer, we drive a condensation of the bound state of the emergent fermion in the $\Z_2$ gauge theory and the $\psi$ in Spin$(2n)_1$. In other words, the $\psi$ in Spin$(2n)_1$ layers are all identified with the fermion in the $\Z_2$ gauge theory. Consequently, $v$ or $v'$ anyons are attached to the $\Z_2$ flux lines. There are no separate anyons confined on the Spin$(2n)_1$ layers anymore, so these layers become invertible defects embedded in the fermionic $\Z_2$ gauge theory. These defects are precisely the TSC defects introduced in the previous paragraph, since the Spin$(2n)_1$ can be thought of as coupling a TSC of Chern number $2n$ to a $\Z_2$ gauge field, and what we just did is to ``Higgs" the emergent $\Z_2$ gauge field in Spin$(2n)_1$ with that of the \spd{3} gauge theory. Notice that the kind of condensation transitions on the Spin$(2n)_1$ layers can be driven by interactions that preserve the $C_r$ symmetry, and it is expected that there is no spontaneous symmetry breaking. As a result, the new $\Z_2$ gauge theory obtained this way still has the $C_r$ symmetry with the same anomaly.

\section{Discussions and conclusions}

In this work we have studied symmetry-preserving gapped boundary states for \spd{4} BC SPT phases protected by $\Z_N$ and $C_N$ symmetries. We show that for $N\notin \{2,4,8,16\}$, no such boundary states exist for the root $\Z_N$ SPT phases. We then propose that for $N=2,4,8,16$ a candidate boundary topological order is a fermionic $\Z_2$ gauge theory, where the anomalous symmetry is generated by topological superconductor defects. We provide explicit constructions of the boundary theory for $C_N$ SPT phases for $N=2,4,8,16$.

One immediate question left open from our analysis is the 32-gappability of Spin$(2n+1)_{2n+1}$. Given that the simplest of the series, SU(2)$_6$, already has 7 anyon types, it is challenging to classify the Lagrangian algebras in SU(2)$_6^{\boxtimes 32}$. We conjecture that Spin$(2n+1)_{2n+1}$ is not 32-gappable.

An interesting question for future works is to construct possible gapless boundary theories for general $N$, or even the $\U$ symmetry group. For $N=2$, a gapless boundary theory was constructed in Ref. [\onlinecite{HuangPRR2020}]. 

It will also be interesting to clarify the relation between the bosonic $\Z_N$ (or $C_N$) SPT phases and the fermionic ones. The fermionic phases can be realized by non-interacting fermions and the natural boundary states are Weyl fermions. One can imagine that certain fermionic phases are actually adiabatically connected to a bosonic one with trivial gapped fermions. In fact, this provides a possible route to construct gapless boundary states for many values of $N$ if the bosonic phase can be ``embedded" into a non-interacting fermionic one. This is the case for all odd $N$, so a possible boundary theory is obtained by gauging fermion parity in a \spd{3} Weyl fermion.

We have discussed the 0-form symmetry in a fermionic $\Z_2$ gauge theory in \spd{3}. The full symmetry group includes $\Z_2$ 1-form and 2-form symmetries, and together with the $\Z_{16}$ 0-form symmetry they are expected to form a 3-group~\cite{Barkeshli:2022wuz}. The 1- and 2-form symmetries and their anomalies are analyzed in Ref. [\onlinecite{Barkeshli:2022edm}]. It is important to fully understand the structure and the anomaly of the 3-group. In addition, one can also consider non-invertible defects to get an even richer structure (conjecturally a fusion 3-category).

\section{Acknowledgement}

M.C. thanks Z. Bi, Y.A. Chen, T. Ellison and S.-J. Huang for helpful conversations. M.C. is particulrly grateful for C.-M. Jian for discussions on the effective response action of \spd{4} BC SPTs, and Juven Wang on the group structure of the classification for $\Z_3$ symmetry. M.C. would like to acknowledge NSF for support under award number DMR-1846109.

\appendix

\section{Group structure of $C_N$ SPT phases}
\label{group-structure}

In the following we denote the root BC SPT phase (with a single $E_8$ state on the rotation center) as $x$. Similarly, denote the root group-cohomology SPT phase (with the root group-cohomology 2D $\Z_N$ SPT state on the center) by $y$. Stacking of phases is denoted additively and $0$ represents the trivial phase. We then have $Ny=0$.

To determine the group structure of $C_N$ SPT phases, we observe that in the block construction~\cite{SongPRX2017}, a state $\ket{\psi}$ with $N$ copies of $E_8$ states, all parallel to the rotation center and arranged in a $C_N$-symmetric configuration is actually adiabatically connected to a trivial state. On the \spd{3} boundary, we have $N$ copies of $(E_8)_1$ CFTs, where the $C_N$ acts as $\Z_N$ cyclic permutations. Apparently, it has the same mixed-gravitational anomaly as the boundary of the $Nx$ phase. On the other hand, it can also have a pure $\Z_N$ anomaly. Recall that 't Hooft anomalies of a $\Z_N$ symmetry are classified by $\H^3[\Z_N, \U]=\Z_N$, so they can be labeled by an integer $\omega\in \Z/N\Z$.  It is known that the $\Z_N$ anomaly $\omega_N$ for $N$ copies of $(E_8)_1$ CFTs is given by
\cite{Bischoff:2018loc}
\begin{equation}
    \omega_N = 
    \begin{cases}
    0 & 3\nmid N\\
    \frac{N}{3} & 3\mid N
    \end{cases}.
\end{equation}
Therefore, the triviality of $\ket{\psi}$ implies $Nx+\omega_Ny=0$ mod $N$.

 For $3\nmid N$, we have $Nx=0$, so the group is $\Z_N^2$. 
 
 For $3\mid N$, we find $Nx+\frac{N}{3}y=0$ mod $N$, so it follows  that  $x$ generates a $\Z_{3N}$ subgroup. The group structure is $\Z_{3N}\times \Z_{N/3}$, where $\Z_{N/3}$ is generated by $3x+y$.

Special cases of the classification (for $N=2,3,4,8$) have been obtained in Ref. [\onlinecite{Wan:2018bns}].

Let us now consider the boundary states for the $N=3^n$ SPTs. Since it is well-known that the group-cohomology SPT phases admit topological boundary theories, for this purpose we mod out the group-cohomology phases from the classification. The remaining group $\Z_{N}$ is generated by the root BC phase $x$. For the $kx$ state, there are $k$ copies of chiral $(E_8)_1$ CFTs at the rotation axis on the \spd{3} boundary. To apply our construction, the \spd{2} MTC needs to have chiral central charge $c_-=\frac{8k}{N}$. Let us write $k=b \cdot 3^a$, where $b$ is the coprime of $3^n$ and $0 \leq a < n$. Therefore, the MTC must have order $3^{n-a}$ in the Witt group, which is impossible. So we conclude that our construction does not work for these SPTs.

\section{U(1) and $\Z_N$ SPT phases in \spd{4}}
\label{Z3}

In this appendix, we discuss in details the computation of the partition functions of $\Z_N$ SPT phases coupled to background gauge field, in particular for the BC phases. For $3\nmid N$, the action given in Eq. \eqref{Z_ZN} correctly describes the BC SPT phases, and evaluating the partition function on $K_3\times S^1$ yields. However, when $3|N$, Eq. \eqref{Z_ZN} does not apply anymore. For example, for $N=3$,
 it is known that for a $\Z/3\Z$ gauge field $A$, $\int A\cup p_1\equiv 0$ mod 3 on any closed 5-manifold \cite{df69ae3e-bbcd-3b86-9506-70e547ea5906}.

 The cobordism classification gives a $\Z_9$ classification for $\Z_3$  SPT phases in \spd{4}, where the generator is the root BC SPT phase. This is consistent with the $C_3$ analysis done in Appendix \ref{group-structure}. We now discuss how to compute the partition function on $K_3\times S^1$ for the $\Z_3$ root BC phase.
 
 In the following we write the partition function as ${\cal Z}=e^{2\pi iS}$, where $S$ is the action. 

\subsection{Topological responses}
To do this, we start from SPT phases with $\U$ symmetry, and then break the symmetry down to $\Z_N$. For a background U(1) gauge field $\tilde{A}$, one can write down the following two topological terms in 5d:
\begin{equation}
    S_1=\int_{B_6} c_1^3, S_2=\int_{B_6}c_1\wedge p_1.
\end{equation}
Here $B_6$ is a 6-dimensional extension of the 5-manifold, and $c_1=\frac{\tilde{F}}{2\pi}$ is the first Chern class. Both terms are well-defined since on closed 6-manifolds they are quantized to integers. 

And yet, the following combination is also integral on a closed 6-manifold $B_6$ based on the Hirzebruch signature theorem:~\footnote{We thank Juven Wang for pointing this out to us.}:
\begin{equation}
    S_3=\int_{B_6} \frac13 \left(c_1^3-c_1\wedge p_1\right).
\end{equation}
As a result, there is a well-defined 5d topological term when $S_3$ is defined on a 6d manifold with boundary. We will schematically write it as
\begin{equation}
    S_3=\int_{M_5}\frac13 \left(\frac{\tilde{A}\wedge \tilde{F}\wedge \tilde{F}}{(2\pi)^2}-\tilde{A}\wedge p_1\right).
\end{equation}

Let's first consider dimensional reduction of both terms $S_1$ and $S_2$ on $S^2\times M_3$, where $M_3$ is a 3-manifold, with a unit flux through $S^2$ (i.e. $\int_{S^2}\frac{\tilde{F}}{2\pi}=1$). For $S_1$, one finds a 3d Chern-Simons term at level $6$, i.e. a bosonic U(1) SPT phase with Hall conductance $\sigma_H=6$. Notice that the most fundamental bosonic U(1) SPT phase has $\sigma_H=2$. For $S_2$, as discussed in Sec. \ref{domain wall} the dimensional reduction gives an invertible bosonic topological phase with chiral central charge $c_-=24$. 

For $S_2$, it is also useful to consider dimensional reduction on $S^1\times M_4$, which can be interpreted as quantization of the theory on a closed spatial manifold $M_4$. We find that the state on $M_4$ carries a $\U$ charge $p_1(M_4)=3\sigma(M_4)$, where $\sigma(M_4)$ is the signature of the manifold. On the other hand, $S_1$ the action does not contribute to the ground state charge on $M_4$. 

Combining $S_1$ and $S_2$, $S_3$ when dimensionally reduced on $S^2\times M_3$ yields a (2+1)d invertible phase with $\sigma_H=2$ and $c_-=-8$, while carries U(1) charge $-\sigma(M_4)$ upon quantization on $M_4$. We can choose $S_3$ and $S_1$ (or $S_2$) as representing the two generators of the U(1) SPT phases in \spd{4}.

\subsection{Defect decorations}
This dimensional reduction procedure is closely related to the decorated defect picture explained in Sec.\ref{domain wall}. Here, the topological defect to consider is a fundamental vortex, which has spatial dimension 2, i.e. a vortex sheet. Note that a vortex sheet is only available when the U(1) symmetry is spontaneously broken. Naively, it might appear that one can not discuss any SPT deocration protected by the original $\U$ symmetry. However, one can combine U(1) with a spatial SO(2) rotation to get a new $\U'$ symmetry, which is respected by the vortex system.   
We can then ask what kind of \spd{2} $\U'$ SRE phase is decorated on the domain wall. It is well-known that such SRE phases are labeled by a pair of integers $[\sigma_H/2, c_-/8]$, where $\sigma_H$ is the Hall conductance and $c_-$ is the chiral central charge ($c_-/8$ gives the number of $E_8$ states). 

To make connection with the topological responses in the previous section, observe that the ``boundary" of a vortex sheet is a U(1) vortex in the symmetry-breaking phase. If one restores the U(1) symmetry (and the SO(2) spatial symmetry as well), the vortex sheet boundary should become the U(1) monopole, and the $\U'$ symmetry is identified with the original $\U$. Therefore, the decoration on the vortex sheet should be identified with the theory obtained from the compactification on $S^2$. We thus have the following identification: $S_3\sim [1,-1], S_1\sim [3,0], S_2\sim [0,3]$. A general topological term $aS_1+bS_3$ should give decoration $[3a+b,-b]$. In other words, there appears to be a constraint 
\begin{equation}
\frac{\sigma_H}{2}+\frac{c_-}{8}\equiv 0\,(\text{mod }3).
\label{eqn:U1constraint}
\end{equation}

 When the symmetry is broken down to $\Z_N$, the decoration on the vortex sheet becomes the one at the junction of $N$ fundamental $\Z_N$ defects (each carrying $2\pi/N$ flux). Equivalently, the decoration is the fusion outcome of $N$ such defects as indicated in Eqn.\ref{nab}. In the symmetry breaking phase, the defects become domain walls, then the configuration of $N$ domain walls meeting at the fusion junction preserves the combined symmetry $\Z_N'$ of the $\Z_N$ and $C_N$ transformation. Then one can ask what kind of \spd{2} $\Z_N'$ SPT phase is decorated on the junction. Recall the discussion in Sec.\ref{domain wall}, such SPT phases are classified by $[k, c_-/8]$, where $k\in \Z/N\Z$ labels the group-cohomology $\Z_N$ SPT phases. 
 
 Notice that there is an additional equivalence relation: one can attach the trivial configuration, $N$ copies of $E_8$ states, to $N$ of $\Z_N$ codimension-1 domain walls in a $C_N$ (thus also $\Z_N'$) symmetric way, increasing $c_-/8$ by $N$. However, as shown in Appendix \ref{group-structure}, when $3\mid N$ the index $k$ is also modified to $k+N/3$. So for $3\mid N$ we find the important equivalence relation
\begin{equation}
 [k, c_-/8]\sim [k+N/3, c_-/8+N].
\end{equation} 
As a result, we have $[0, N]\sim [-N/3,0]$, and hence $[0,1]$ generates a $\Z_{3N}$ subgroup. The other $\Z_{N/3}$ subgroup is generated by $[1,3]$.

For $3\nmid N$ we instead have $[k, c_-/8]\sim [k, c_-/8+N]$, and the classification is $\Z_N\times\Z_N$, with the two generators being $[k=1,c_-=0]$ and $[k=0, c_-/8=1]$.

If the phase can be connected to a U(1) SPT phase, then we have $k=\sigma_H/2$ mod $N$ satisfying the constraint Eq.\eqref{eqn:U1constraint} for U(1) SPT phase. Therefore, when $3\mid N$ certain $\Z_N$ group-cohomology SPT states, e.g. $[k=1,c_-=0]$ for $N=3$, can not be lifted to a U(1) SPT phase. The topological response for such phases cannot be described by $S_1$, instead we define $S_1'$ as the topological response for the $[k=1,c_-=0]$ phase. Note that for $3\nmid N$, one can easily show that all $\Z_N$ SPT phases can be embeded to U(1) SPTs.

From this analysis, the correspondence between the $C_N$ and $\Z_N$ classifications can also be made clear. In fact, the decoration $[\sigma_H/2, c_-/8]$ in the $\Z_N$ case is precisely the SRE state placed on the rotation center in the $C_N$ block construction.

\subsection{Computing $Z(K_3\times S^1)$}
For $\Z_N$ symmetry, if there is a nontrivial holonomy $e^{2\pi i/N}$ along $S^1$, $Z(K_3\times S^1)$ is given by $e^{2\pi i q(K_3)/N}$, where $q(K_3)$ is the $\Z_N$ charge of the state on $K_3$. The group-cohomology SPT phases always give $q(K_3)=0$ (at least within TQFT), thus only the BC SPT phases contribute nontrivially to the charge $q(K_3)$. 

For $3\nmid N$, we can always embed the $\Z_N$ SPT phase to a $\U$ SPT phase, and it follows from the dimensional reduction that $q(K_3)=-\sigma(K_3)\frac{c_-}{8}=-2c_-$ mod $N$. When $N=3$, $q(K_3)=-\sigma(K_3)=-16 \equiv 2$ mod $3$, corresponding to the $\Z_{3N}$ subgroup generated by $[0,1]$. Thus the partition function on $K_3\times S^1$ is $e^{4 \pi i/3} \neq 1$.

Now let us consider $N=3^p$ symmetry, with $p>1$. Again, since group-cohomology SPT phases do not contribute to $q(K_3)$, we can mod out the $\Z_{3N}\times \Z_{N/3}$ group by the $\Z_N$ group generated by the group-cohomology SPTs, leaving only the $\Z_N$ BC SPTs. Essentially, the quotient is given by the mapping $[k, c_-/8]\Rightarrow c_-/8$ mod $N$. For the purpose of computing $q(K_3)$, one can use $S_3$ (with the symmetry reduced from U(1) to $\Z_N$) for the generator of this quotient group, which gives $q(K_3)=-\sigma(K_3)=-16$.  Note that the generator of $\Z_{N/3}$ corresponds to $3$ after quotient, so the partition function on $K_3\times S^1$ is $e^{6\pi iq(K_3)/N}=e^{-32\pi i/3^{p-1}}\neq 1$ as long as $p>1$.


In [\onlinecite{Wan:2018bns}], it was claimed that the response action for the root BC phase should be the Postnikov square: $\beta_{(9,3)}(\beta_{(3,3)} A\cup \beta_{(3,3)}A)$, where $A$ is the $\Z_3$ gauge field and $\beta_{(n,m)}$ is the Bockstein homomorphism associated with the extension $\Z_n \xrightarrow []{\cdot m}\Z_{nm} \rightarrow \Z_{m}$. It will be interesting to understand the relation between the various forms of actions.

\section{Witt group of MTCs}
\label{app:witt}

In this section we will review the definition of the Witt group, as well as some known facts about it.

First we define the notion of Witt equivalence between two \spd{2} topological phases. Two topological phases $\cal{B}_1$ and $\cal{B}_2$ are Witt equivalent, if $\cal{B}_1\boxtimes \ol{\cal{B}}_2$ has a fully gapped interface to an invertible state (i.e. some copies of $E_8$ states). Here $\ol{\cal{B}}_2$ is the mirror image of $\cal{B}_2$. In other words, there is a gapped interface between $\cal{B}_1$ and $\cal{B}_2$ as long as we are allowed to freely stack copies of $E_8$ states. Mathematically, two MTCs $\cal{B}_1$ and $\cal{B}_2$ are Witt equivalent if $\cal{B}_1\boxtimes \ol{\cal{B}}_2$ is a quantum double (Drinfeld center of some fusion category).

Below we review known results about the torsion subgroup of the Witt group, particularly for Abelian MTCs. We adopt notations in Ref. [\onlinecite{Bonderson2012}] for MTCs. In particularly, $\Z_N^{(k)}$ refers to an Abelian topological order with $N$ Abelian anyons labeled by $[a]_N$, where $a\in \{0,1,\dots, N-1\}$, and $[\cdot]_N$ denotes mod $N$. The fusion rules are given by addition: $[a]_N\times [b]=[a+b]_N$. The topological twist factor $\theta_{[a]_N}=e^{i\frac{\pi k}{N} a^2}$. Notice that for odd $N$, $k$ must be an integer. While for even $N$, $k$ can be an integer or half-integer, but only half-integer values give modular theories. $\Z_{2n}^{(1)}$ represents the topological order of the familiar $\U_{2n}$ Chern-Simons theories.

 The Witt group of all Abelian MTCs, denoted by $\mathcal{W}_\text{pt}$ following the notation in Ref. [\onlinecite{DGNOWitt2}], has the following decomposition:
\begin{equation}
    \mathcal{W}_\text{pt}=\bigoplus_{p\text{ prime}}\mathcal{W}_\text{pt}(p).
\end{equation}
For each prime $p$, the $p$-subgroup $\mathcal{W}_\text{pt}(p)$ is given by:

\begin{description}
\item[$p=2$]: $\mathcal{W}_\text{pt}(2)=\Z_8\times \Z_2$. Here $\Z_8$ is generated by the semion theory $\Z_2^{(1/2)}$, and $\Z_2$ is generated by $\Z_2^{(1/2)}\times\ol{\Z_4^{(1/2)}}$.
\item[$p\equiv 1\,(\text{mod }4)$]: $\mathcal{W}_\text{pt}(p)=\Z_2\times \Z_2$. One generator can be chosen as $\Z_p^{(1)}$, and the other $\Z_p^{(k)}$ where $k$ is a quadratic non-residue mod $p$.
\item[$p\equiv 3\,(\text{mod }4)$]: $\mathcal{W}_\text{pt}(p)=\Z_4$. The generator could be any $\Z_p^{(n)}$ theory for $1\leq n<p$.
\end{description}
 Another  important example is Kitaev's 16-fold way: the Ising MTC generates a $\Z_{16}$ group, which contains order-$2,4,8$ subgroups. 

 There also exists infinitely many order-$32$ elements in the Witt group. They are represented by the Spin$(2n+1)_{2n+1}$ Chern-Simons theories, where $n\geq 1$. The simplest one of them is $\text{Spin}(3)_3\simeq \text{SU}(2)_6$. It is the ``square root" of an Ising Witt class: two copies of Spin$(2n+1)_{2n+1}$ is Witt equivalent to Spin($(2n+1)^2)_1$.

 We will now examine the $N$-gappability of the examples mentioned above.

\subsection{$\Z_p^{(n)}$}

We start from the $\Z_p^{(n)}$ theories, where $p$ is an odd prime.

First, we consider $p\equiv 3$ mod $4$, and $\Z_p^{(n)}$ MTCs have order 4 in the Witt group. We now show that they are not $4$-gappable.

 Label anyon in four copies of $\Z_p^{(n)}$ by $a=(a_1,a_2,a_3,a_4)$, where $a_i\in\{0,1,\dots, p-1\}$. The most general form of a cyclic permutation is generated by 
\begin{equation}
   g(a)=(s_1a_4,s_2 a_1,s_3a_2,s_4a_3), 
\end{equation} 
 where $s_i=\pm 1$. Basically, this is a ``bare" permutation that takes $a$ to $(a_4,a_1,a_2,a_3)$, combined with a topological symmetry of each of the $\Z_p^{(n)}$ layer. It is known that the only nontrivial topological symmetry of the $\Z_p^{(n)}$ MTC is the charge conjugation $a\rightarrow -a$.

 In our setup, we require $g^4$ is the identity. 
 Under repeated actions of $g$ we have:
\begin{equation}
\begin{split}
    (a_1,a_2,a_3,a_4)&\rightarrow (s_1a_4,s_2a_1,s_3a_2,s_4a_3)\\
    &\rightarrow (s_1s_4a_3,s_2s_1a_4,s_3s_2a_1,s_4s_3a_2)\\
   & \rightarrow
    (s_1s_4s_3a_2, s_2s_1s_4a_3,s_3s_2s_1a_4,s_4s_3s_2a_1)\\
    &\rightarrow
    s_1s_2s_4s_4(a_1, a_2,a_3,a_4).
\end{split}
\end{equation}
So we must have $s_1s_2s_3s_4=1$, otherwise $g^4$ is the global charge conjugation, which acts nontrivially in $\Z_p$ theories.

If $a$ is in the Lagrangian subgroup and the subgroup preserves the $\Z_4$ symmetry, then $a, g(a), g^2(a), g^3(a)$ must form a condensable subset. Then they must all be bosons, which gives
\begin{equation}
    a_1^2+a_2^2+a_3^2+a_4^2\equiv 0\ \text{mod }p.
\end{equation}
In addition, they must have trivial mutual braiding statistics, which lead to
\begin{equation}
\begin{gathered}
s_2a_1a_2+s_3a_2a_3+s_4a_3a_4+s_1a_4a_1\equiv 0\ \text{mod }p\\
    (s_1s_4+s_2s_3)a_1a_3+(s_1s_2+s_3s_4)a_2a_4\equiv 0\ \text{mod }p. 
\end{gathered}
\end{equation}
The last equation simplifies to $2(s_1s_4a_1a_3+s_1s_2a_2a_4)\equiv 0$ mod $p$, and since $2$ is invertible mod $p$, we have 
\begin{equation}
(s_1s_4a_1a_3+s_1s_2a_2a_4)\equiv 0\ \text{mod }p.
\end{equation}
Since $s_1s_2s_3s_4=1$, we 
 can represent $s_1=c_4c_1,s_2=c_1c_2,s_3=c_2c_3,s_4=c_3c_4$, and define $b_i=c_ia_i$. The equations are simplified to
\begin{equation}
    \begin{gathered}
        b_1^2+b_2^2+b_3^2+b_4^2\equiv 0\ \text{mod }p,\\ b_1b_2+b_2b_3+b_3b_4+b_4b_1\equiv 0\ \text{mod }p,\\ 
        b_1b_3+b_2b_4\equiv 0\ \text{mod }p.
    \end{gathered}
    \label{eqn:relations_b}
\end{equation}
Together they imply $(b_1+b_2+b_3+b_4)^2\equiv 0$ mod $p$. Then the second relation leads to {
\begin{equation*}
\begin{split}
b_1b_2+b_2b_3+b_3b_4+b_4b_1&=(b_2+b_4)(b_1+b_3)\\
&\equiv-(b_1+b_3)^2\equiv 0\ \text{mod }p,
\end{split}
\end{equation*}
}
from which we conclude that $b_3\equiv -b_1, b_4\equiv -b_2$ mod $p$. The last relation in Eq. \eqref{eqn:relations_b} then implies $-2(b_1^2+b_2^2)\equiv 0$ mod $p$. For $p$ odd, we then have $b_1^2+b_2^2\equiv 0$ mod $p$.

One can show that there is no solution for $p\equiv 3\ (\text{mod }4)$. To show this is the case, we write $b_1^2\equiv q$, so $b_2^2\equiv -q$. In other words, both $q$ and $-q$ are quadratic residues of $p$. Compute the Legendre symbols:
\begin{equation}
    \left(\frac{-q}{p}\right)=\left(\frac{-1}{p}\right)\left(\frac{q}{p}\right)=(-1)^{\frac{p-1}{2}}\left(\frac{q}{p}\right)=-\left(\frac{q}{p}\right).
\end{equation}
So it is impossible to have both $q$ and $-q$ being quadratic residues when $p\equiv 3\ (\text{mod }4)$.

Let us now turn to $p\equiv 1\ (\text{mod }4)$, and the $\Z_p^{(n))}$ theories have order 2 in the Witt group. We will show that they are not $2$-gappable.  Using the same argument, $(a_1,a_2)$ and $(s_1a_2,s_2a_1)$ should form a condensable subgroup, which requires
\begin{equation}
    a_1^2+a_2^2\equiv 0\ \text{mod }p, (s_1+s_2)a_1a_2\equiv 0\ \text{mod }p.
\end{equation}
Thus we need to have $s_1=-s_2$. However, under this permutation $(a_1,a_2)\rightarrow (s_1a_2,s_2a_1)\rightarrow s_1s_2(a_1,a_2)$, if $s_1s_2=-1$ then $g^2$ is equal to the charge conjugation. 

\subsection{$\Z_2^{(\frac12)}$}
$\Z_2^{(\frac12)}$ has order 8 in the Witt group. A Lagrangian subgroup in 8 copies of $\Z_2^{(\frac12)}$ should have dimension $16$, so generated by four bosons. 

We first find all bosons, such that the its image under $\Z_8$ form a condensable subgroup. It turns out that there are $8$ such bosons, and they form $\Z_2^3$ group generated by $(1,1,0,0,1,1,0,0),(0,1,1,0,0,1,1,0),(0,0,1,1,0,0,1,1)$. So it is impossible for find a Lagrangian subgroup invariant under $\Z_8$. Interestingly, if we actually condense this $\Z_2^3$ subgroup, we find a $\Z_2$ toric code, and the $\Z_8$ generator acts as electromagnetic duality in this theory. Therefore, we can not further condense bosons without breaking the $\Z_8$ symmetry.

\section{Lattice model realization}

Given the subtleties in the TQFT classification of SPT phases, it is highly desirable to find microscopic constructions of the nontrivial states. For the $N=2$ case, an exactly solvable model was given in [\onlinecite{Fidkowski4DSPT}] (see also [\onlinecite{Chen:2021xks}] for an alternative construction, which applies to $N=4$ as well.). The key observation there is the following: because $p_1\equiv w_2^2\,(\text{mod }2)$, we can interpret the action in Eq. \eqref{Z_ZN} as $\Z_2$ domain walls being decorated with the ``$w_2^2$" SPT states, which are realized as the ground state of the so-called 3-fermion Walker-Wang (WW) model~\cite{BurnellPRB2014}. The key ingredient in this construction is a quantum cellular automata (QCA), or a locality-preserving unitary, that disentangles the 3-fermion WW state. In addition, the QCA exactly squares to $1$. With such a QCA, a wavefunction of a consistent, equal-weight superposition of decorated domain wall states, as well as a commuting projector parent Hamiltonian, can be written down. 

It is not clear whether similar constructions can be generalized to other $N>2$ cases. If we simply generalize the construction in [\onlinecite{Fidkowski4DSPT}], according to the action in \eqref{Z_ZN}, the $\Z_N$ domain wall is decorated by a \spd{3} gapped state, partition function of which is given by $e^{\frac{2\pi i}{N}\int p_1}$. In addition, the disentangling QCA for this state must have order $N$. The constraints on QCAs could be seen from the Witt group of \spd{2} modular tensor categories (MTC), which are mathematical theories describing the universal bulk properties of topological phases. While a complete topological classification of QCAs in \spd{3} is still unknown, there is a growing body of evidences~\cite{Haah:2018jdf, Haah:2019fqd, Shirley:2022lhu} suggesting that they are classified by the Witt group of \spd{2} MTCs. We review the definition of Witt group in Appendix \ref{app:witt}. Conjecturally, a QCA that disentangles a WW model with the input MTC in a nontrivial Witt class is topologically nontrivial. The partition function for the WW model is $e^{\frac{2\pi ic_-}{24}\cdot p_1}$, where $c_-$ is the chiral central charge of the input MTC. Assuming that this conjectured classification of QCA is correct, we conclude that in order to generalize the construction to $\Z_N$, we would need to find a topological phase with chiral central charge $\frac{48}{N}$ mod 8 and with order $N$ in the Witt group.

However, as already mentioned in Section \ref{constraint}, by the constraint of the order of elements in the Witt group. The only possible finite values for $N$ are $2^n$ with $1\leq n\leq 5$ (there are obviously elements of infinite order). This immediately rules out any odd $N>3$ in the construction.

Even for $N$ that divides 48, when there exist order-$N$ elements in the Witt group, there is a further constraint.  If the MTC has an order $N$ element in the Witt group, the corresponding QCA conjecturally is also of order $N$, meaning that $N$-th power of the QCA is a finite-depth local unitary circuit. However, for the construction to work, the $N$-th power needs to be exactly $1$. So far this has only been done for the 3-fermion QCA with $N=2$, and to the best of our knowledge, no other known examples of QCA satisfy this property. 

\section{Algebraic description of symmetry-preserving anyon condensation}
\label{app:condensation}
We review the algebraic theory of gapped boundaries of a two-dimensional topological phase~\cite{kong2014, kitaev2012, eliens2013, lan2015, NeupertPRB2016, Cong2017}, closely following the formulation in [\onlinecite{Cong2017}] and [\onlinecite{ChengPRR2020}]. We extensively use the language of unitary modular tensor category (UMTC) for \spd{2} topological phases. A brief summary of UMTC in this context can be found in the appendix of [\onlinecite{ChengPRR2020}]. 

	A gapped boundary corresponds to a Lagrangian algebra of the bulk MTC. Physically the Lagrangian algebra indicates which bulk anyons are condensed on the boundary~\cite{kong2014, levin2013}.
	
The mathematical theory of the gapped boundary takes into account the local process of annihilating a condensable anyon $a$ on the boundary. Similar to fusion/splitting spaces, we associate a vector space for local operators that annihilate $a$, denoted as $V^a$, with basis vector $\ket{a;\mu}$.  The dimension of this vector space is the ``multiplicity'' $n_a$ of $a$ in the Lagrangian algebra. Obviously we must have $n_1=1$. 
 
Diagrammatically, the condensation process is represented by an anyon line terminating on a wall representing the boundary. We also attach a label at the termination point which represents the state of the boundary condensation space. When $n_a=1$ it can be suppressed.

\begin{figure}
    \centering
    \includegraphics[width=0.65\columnwidth]{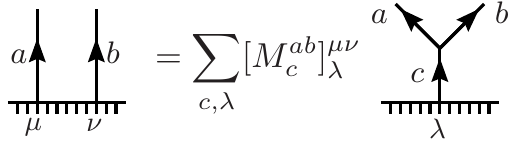}
    \caption{A diagrammatic representation of the M symbol.}
    \label{fig:M}
\end{figure}
An important property of the algebra is the following ``$M$ symbol'':
\begin{equation}
    \ket{a;\mu}\ket{b;\nu} = \sum_{c,\lambda}[M^{ab}_c]^{\mu\nu}_{\lambda}\ket{c;\lambda}.
\end{equation}
The definition is illustrated diagrammatically in Fig. \ref{fig:M}.

Next we impose consistency conditions on the $M$ symbols. We can apply $M$ moves to three anyon lines terminating $a,b,c$ on the boundary, but in different orders, which leads to a variation of the pentagon equation: 
\begin{equation}
	\sum_{e,\sigma}[M^{ab}_e]^{\mu\nu}_\sigma [M^{ec}_d]^{\sigma\lambda}_{\delta}[F^{abc}_d]_{ef}=\sum_{\psi}[M^{af}_{d}]^{\mu\psi}_\delta [M^{bc}_f]^{\nu\lambda}_\psi 
	\label{eqn:Mfusion}
\end{equation}

The $M$ symbols also have gauge degrees of freedom, originating from the basis transformation of the boundary condensation space $V^a$: $\widetilde{\ket{a;\mu}}=\Gamma^a_{\mu\nu}\ket{a;\nu}$, where $\Gamma^a_{\mu\nu}$ is a unitary transformation. The $M$ symbol becomes
\begin{equation}
	[\tilde{M}^{ab}_c]^{\mu\nu}_\lambda=\sum_{\mu',\nu',\lambda'}\Gamma^{a}_{\mu\mu'}\Gamma^b_{\nu\nu'}[M^{ab}_c]^{\mu'\nu'}_{\lambda'}[\Gamma^c]_{\lambda'\lambda}^{-1}.
	\label{}
\end{equation}
$M$ symbols are affected by the gauge transformation of bulk fusion space as well.

It is convenient to fix the gauge for the following symbols:
\begin{equation}
	[M^{1a}_a]^{\mu}_\nu=[M^{a1}_a]^{\mu}_\nu=\delta_{\mu\nu}.
	\label{}
\end{equation}

Braiding puts further constraints on the $M$ symbols. Since the anyons condense on the boundary, it should not matter in which order the anyon lines terminate on the boundary. 
\begin{equation}
	[M^{ba}_c]^{\nu\mu}_\lambda R^{ab}_c=[M^{ab}_c]^{\mu\nu}_\lambda.
	\label{eqn:Mbraiding}
\end{equation}
There is a similar condition for the inverse braiding.

It was shown in Ref.~[\onlinecite{Cong2017}] that these conditions are equivalent to the mathematical definition of a commutative, connected and separable Frobenius algebra $\mathcal{A}=\bigoplus_a n_a a$ in a braided tensor category,  with the algebra morphism $\mathcal{A}\times \mathcal{A}\rightarrow \mathcal{A}$ precisely given by the $M$ symbol. 

\subsection{Symmetry-preserving condensation}
\label{sec:charged-condensate}
We now give a precise definition of anyon condensation that preserves the global symmetry~\onlinecite{Bischoff2019}.  This builds on top of the algebraic theory of \spd{2} symmetry-enriched topological phases, known as the $G$-graded braided tensor category, reviewed in [\onlinecite{ChengPRR2020}]. For a more complete account see [\onlinecite{SET}]. A key fact that we will use is that in a topological phase enriched by symmetry group $G$, the symmetry action on anyons is fully specified by the following data: $\rho_\mb{g}, U_\mb{g}(a,b;c)$ and $\eta_a(\mb{g,h})$. Here $\rho_\mb{g}$ denotes a permutation of anyon labels, and in the following we write ${}^\mb{g}a\equiv \rho_\mb{g}(a)$. $U_\mb{g}(a,b;c)$ are unitary transformations acting on the fusion spaces. $\eta_a(\mb{g,h})$ are phase factors that describe projective symmetry transformations on individual anyons. $\rho, U$ and $\eta$ need to satisfy consistency conditions given in [\onlinecite{SET}]. Again we refer to [\onlinecite{SET}] for their definitions and properties. In the present case of the $\cal{B}^{\boxtimes N}$ MTC, $\rho$ is the $\Z_N$ cyclic permutation, and $U$ and $\eta$ are both $1$ if we use the natural gauge choice where the $F$ and $R$ symbols of $\cal{B}^{\boxtimes N}$ are simply given by the Cartesian products of those of $\cal{B}$.

In the following $a,b,c, \dots$ denote anyons in the condensate, unless otherwise specified. We assume $n_a=1$ whenever $a$ belongs to the condensate, so we omit the index for the boundary condensation space.
   Since the boundary is fully gapped and symmetric, we can posit that for each $\mb{g}$ there exists at least one $\mb{g}$-defect that can be absorbed  without creating any additional excitations on the boundary.

\begin{figure}
    \centering
    \includegraphics[width=0.65\columnwidth]{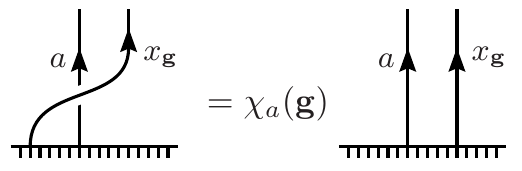}
    \caption{A diagrammatic representation of the $\chi$ symbol.}
    \label{fig:chi}
\end{figure}

Here $\chi_a(\mb{g})$ is a phase factor. Physically, $\chi_a(\mb{g})$ encodes the $\mb{g}$ action on the condensed anyon, illustrated in Fig.\ref{fig:chi}. When there is more than one condensation channel, $\chi$ should be replaced by a unitary transformation acting on the condensation space. 

If we slide a vertex which splits a $\mb{gh}$-defect to $\mb{g}$- and $\mb{h}$-defects over a boundary vertex, we find
\begin{equation}
	\eta_{a}(\mb{g,h})=\frac{\chi_a(\mb{gh})}{\chi_{ {}^{\bar{\mb{g}}}a}(\mb{h})\chi_a(\mb{g})}.
	\label{eqn:defect-bdr1}
\end{equation}

We can also consider fusion of condensable anyons on a boundary. For $a,b,c$ in the condensate, sliding a $\mb{g}$ line over the diagrammatic equation that defines $M$ symbol, one finds
\begin{equation}
	M^{{}^{\bar{\mb{g}}}a, {}^{\bar{\mb{g}}}b}_{{}^{\bar{\mb{g}}}c}U_\mb{g}(a,b;c)= M^{ab}_c \frac{\chi_a(\mb{g})\chi_b(\mb{g})}{\chi_c(\mb{g})}.
	\label{eqn:defect-bdr2}
\end{equation}

We believe that these two conditions Eqs.~\eqref{eqn:defect-bdr1} and~\eqref{eqn:defect-bdr2} are sufficient and necessary for the condensation to preserve symmetry.  Mathematically, $\chi_a(\mb{g})$ defines an algebraic isomorphism for each $\mb{g}$. The consistency conditions guarantee that one has a $G$-equivariant algebra structure on $\cal{L}$~\cite{Bischoff2019}.

Note that for a $\mb{g}$-invariant anyon $a$, $\chi_a(\mb{g})$ can be interpreted as the $\mb{g}$ charge carried by $a$. In particular, it means that $\eta_a(\mb{g,h})=\frac{\chi_a(\mb{gh})}{\chi_a(\mb{g})\chi_a(\mb{h})}$. Below, in all our examples the bulk has no symmetry fractionalization so we make the canonical choice that $\eta_a(\mb{g,h})=1$.  

\subsection{Condensation in Spin$(2n)_1$ layers}\label{M:general_solution}
Below we focus on the cases of Spin$(2n)_1^{\boxtimes r}$. We will only study the $n=1,2$ cases. $n=4$ has been treated in Ref. [\onlinecite{Qi:2017qui}], and the other values of $n$ are similar. We start from a few general results that apply to all $n$. As before, anyons are denoted by a $r$-tuple.

Denote the Lagragian subgroup as $\cal{A}$ for the system, which has two types of group elements. The subgroup $\cal{A}_0$ consists of all bosons $\{\mb{a}_0,\mb{b}_0, \dots \} \in \cal{A}_0$ made of an even number of fermions while the rest of the group elements are expressed as $\mb{v}+\mb{a}_0$, where $\mb{v}$ is the fermion parity flux. 

We observe that $F^{\mb{a}_0,\mb{b}_0,\mb{c}_0}=1$. So $M^{\mb{a}_0,\mb{b}_0}$ forms a 2-cocycle over $\cal{A}_0$. We will postulate that $M^{\mb{a}_0,\mb{b}_0}=1$ in our solutions.

To simplify Eqn. \ref{eqn:Mfusion}, we make the following gauge choices:
By using the $\Gamma^{\mb{v}+\mb{a}_0}$ gauge freedoms, we set $M^{\mb{v},\mb{a}_0}=1$. Then using $\Gamma^\mb{v}$ we set $M^{\mb{v},\mb{v}}=1$. From the consistency equations we find the following expressions:
\begin{equation}\label{M_solutions}
\begin{split}
M^{\mb{v}+\mb{a}_0,\mb{b}_0}&=F^{\mb{v},\mb{a}_0,\mb{b}_0},\\ M^{\mb{a}_0,\mb{v}+\mb{b}_0}&=\frac{M^{\mb{a}_0+\mb{b}_0,\mb{v}}}{M^{\mb{b}_0,\mb{v}}},\\
M^{\mb{v}+\mb{a}_0,\mb{v}+\mb{b}_0}&=M^{\mb{a}_0,\mb{v}}.
\end{split}
\end{equation}
Now $M^{\mb{a}_0,\mb{v}}$ has to satisfy
\begin{multline}
M^{2\mb{v}+\mb{a}_0+\mb{b}_0+\mb{c}_0,\mb{v}}M^{\mb{b}_0,\mb{v}}F^{\mb{v}+\mb{a}_0,\mb{v}+\mb{b}_0,\mb{v}+\mb{c}_0}=\\
M^{\mb{a}_0,\mb{v}}M^{\mb{c}_0,\mb{v}}F^{\mb{v},\mb{a}_0,2\mb{v}+\mb{b}_0+\mb{c}_0}.
\label{Mcond2}
\end{multline}

We can write a general solution of the consistency equations $M^{\mb{a,b}}=M_0^{\mb{a,b}}\omega^{\mb{a,b}}$, consisting of a special solution $M_0^{\mb{a},\mb{b}}$ of Eqn. \ref{Mcond2} and a group 2-cocycle $\omega^{\mb{a},\mb{b}}$ over $\cal{A}$. Then $\omega$ will be fixed through Eqn. \ref{eqn:Mbraiding}.



\subsection{$\Z_4$-symmetric Lagrangian algebra of Spin$(4)_1^{\boxtimes 4}$}
We denote the anyons in Spin$(4)_1^{\boxtimes 4}$ by $\mb{a}=(\vec{a}_1,\vec{a}_2,\vec{a}_3,\vec{a}_4)$, where $\vec{a}_i=(a_i^1 \; a_i^2)$ are defined mod 2. The F and R symbols are given by
\begin{equation}
\begin{split}
    F^{\mb{a},\mb{b},\mb{c}} &= \exp \left( \frac{i \pi}{2}\sum_{i=1}^4 \vec{a}_i\cdot(\vec{b}_i+\vec{c}_i-[\vec{b}_i+\vec{c}_i]_2) \right),\\ R^{\mb{a,b}}&=\exp\left[\frac{i\pi}{2} \sum_{i=1}^4\vec{a}_i\cdot\vec{b}_i\right].
\end{split}
\end{equation}
The $\Z_4$ symmetry generator $\mb{g}$ acts on the anyons in the following way:
\begin{equation}
    \rho:(\vec{a}_1,\vec{a}_2,\vec{a}_3,\vec{a}_4)\rightarrow (\vec{a}_2,\vec{a}_3,\vec{a}_4,\vec{a}_1).
\end{equation}
It is obvious that F and R symbols are invariant under $\rho$:
\begin{equation}
    F^{\rho(\mb{a}),\rho(\mb{b}),\rho(\mb{c})}=F^{\mb{a,b,c}}, R^{\rho(\mb{a}),\rho(\mb{b})}=R^{\mb{a,b}}.
\end{equation}
Therefore the corresponding $U$ symbols are all 1. We can then set all the $\eta$ symbols to 1 as well.

It will also be convenient to pick a set of generators for the Lagrangian subgroup. We will use the following set:
\begin{equation}
\begin{split}
    &\mb{v}_1=(1\;0,1\;0,1\;0,1\;0)\\
    &\mb{v}_2=(1\;1,1\;1,0\;0,0\;0)\\
    &\mb{v}_3=(0\;0,1\;1,1\;1,0\;0)\\
    &\mb{v}_4=(0\;0,0\;0,1\;1,1\;1)
    \end{split}
\end{equation}
Together they form a $\cal{A}=\Z_2^4$ group. Then any group element $\mb{a}$ can be expanded in terms of the generators: $\mb{a}=\sum_{i=1}^4 \tilde{a}_i\mb{v}_i$.

Following the procedure in Section \ref{M:general_solution}, we find the special solution $M_0^{\mb{a},\mb{b}}$:
\begin{equation}\label{M:Z4_solutions}
\begin{split}
&M_0^{\mb{a}_0,\mb{v}}=e^{\frac{i\pi}{2}\mb{v}\cdot \mb{a}_0},
\end{split}
\end{equation}
and the group 2-cocycle $\omega^{\mb{a},\mb{b}}$:
\begin{equation}
    \omega^{\mb{a,b}}=e^{i \pi (\tilde{a}_2\tilde{b}_3+\tilde{a}_3\tilde{b}_4)}.
\end{equation}

We find the following solution of $\chi$:
\begin{equation}
    \chi_\mb{a}(\mb{g})=e^{i \pi (1-\tilde{a}_4)(\tilde{a}_2+\tilde{a}_3)}.
\end{equation}
and
\begin{equation}
    \chi_{\mb{a}}(\mb{g}^2) =e^{i \pi [(1-\tilde{a}_4)\tilde{a}_3+(1-\tilde{a}_3)\tilde{a}_2]}.
\end{equation}
One can easily verify that all $\mb{g}$-invariant anyons have $\chi=1$. The same is true for all $\mb{g}^2$-invariant anyons.

\subsection{$\Z_8$-symmetric Lagrangian algebra of $\U_4^{\boxtimes 8}$}

\def\ta{\tilde{a}}
\def\tb{\tilde{b}}

We denote the anyons in $\U_4^{\boxtimes 8}$ by a 8-tuple $\mb{a}=(a_1,a_2,\dots,a_8)$, where $a_i\in \{0,1,2,3\}$ are defined mod 4. Since the theory is Abelian, fusion will be denoted additively. We will use short-hand notations for the F and R symbols $(F^{\mb{a,b,c}}_{\mb{a+b+c}})_{\mb{a+b,b+c}}\equiv F^{\mb{a,b,c}}, R^{\mb{a,b}}_{\mb{a+b}}=R^{\mb{a,b}}$. They are given by
\begin{equation}
\begin{split}
    F^{\mb{a,b,c}}&=\exp \left(i\pi \mb{a}\cdot \frac{\mb{b+c-[b+c]_4}}{4}\right), \\
    R^{\mb{a,b}}&=\exp \left(\frac{i\pi}{4}\mb{a\cdot b}\right).
    \end{split}
\end{equation}
The $\Z_8$ symmetry generator $\mb{g}$ acts on the anyons in the following way:
\begin{equation}
    \rho:(a_1,a_2,\dots,a_8)\rightarrow (a_2,a_3,\dots,a_1).
\end{equation}
It is obvious that F and R symbols are invariant under $\rho$:
\begin{equation}
    F^{\rho(\mb{a}),\rho(\mb{b}),\rho(\mb{c})}=F^{\mb{a,b,c}}, R^{\rho(\mb{a}),\rho(\mb{b})}=R^{a,b}.
\end{equation}
Therefore the corresponding $U$ symbols are all 1. We can then set all the $\eta$ symbols to 1 as well.

The Lagrangian subgroup is generated by the following anyons:
\begin{equation}
\begin{split}
    &\mb{v}_1=(1,1,1,1,1,1,1,1)\\
    &\mb{v}_2=(2,2,0,0,0,0,0,0)\\
    &\mb{v}_3=(0,2,2,0,0,0,0,0)\\
    &\mb{v}_4=(0,0,2,2,0,0,0,0)\\
    &\mb{v}_5=(0,0,0,2,2,0,0,0)\\
    &\mb{v}_6=(0,0,0,0,2,2,0,0)\\
    &\mb{v}_7=(0,0,0,0,0,2,2,0).
    \end{split}
\end{equation}
Together they generate a $\mathcal{A}=\Z_2^6\times\Z_4$ group. We find the solution to be 
\begin{equation}\label{M:Z8_solutions}
\begin{split}
&M_0^{\mb{a}_0,\mb{v}}=e^{\frac{i\pi}{4}\mb{v}\cdot \mb{a}_0},\\
\end{split}
\end{equation}
and
\begin{equation}
    \omega^{\mb{a,b}}=e^{i \pi (\tilde{a}_2\tb_3+\ta_3\tb_4+\ta_4\tb_5+\ta_5\tb_6+\ta_6\tb_7)}.
\end{equation}
The solution for $\chi$ is given by
\begin{equation}
    \chi_\mb{a}(\mb{g})=e^{i \pi (1-\ta_7)(\ta_2+\ta_4+\ta_6)}.
\end{equation}
In this case, the only $\mb{g}$-invariant anyons are generated by $\mb{v}_1$, so obviously $\chi_{\mb{v}_1}(\mb{g})=1$. The same is true for other symmetry transformations.

\section{Gauging $\Z_N$ cyclic permutation symmetry}
\label{app:gauging}
We describe the $\Z_N$ gauging of $\cal{B}^{\boxtimes N}$, where the $\Z_N$ generator $g$ acts as
\begin{equation}
    (a_1, a_2,\cdots, a_N)\rightarrow (a_N, a_1, \cdots, a_{N-1}).
\end{equation}
We will denote the gauged theory by $\cal{U}$. Anyons in $\cal{U}$ will be labeled by $(x_{g^r},\chi)$, where $g^r$ is the symmetry flux, and $\chi$ is an irreducible representation of the stabilizer group. Here since $G$ is Abelian we can view $\chi$ as a group homomorphism from the stabilizer group to $\U$. For example, if $r=0$, then $x$ runs through all orbits of anyons $[\mb{b}]$ in $\mathcal{B}^{\boxtimes N}$ under $G$, where  $\mb{b}=(b_1,b_2,\cdots, b_N)$ is a representative element of the orbit. Its stabilizer group, i.e.  the subgroup of $G$ that keeps the anyon $\mb{b}$ invariant, is denoted by $G_{\mb{b}}$. When the stabilizer group is strictly smaller than $G$, we call $([\mb{b}]_1,\chi_\mb{b})$ a superposition anyon.
In general, the label $x$ can actually be chosen as an element in $\cal{B}^{\boxtimes (N,r)}$. 

 Among all $(x_{g^r}, \chi)$ there is one with the minimal quantum dimension, which we will call $(\id_{g^r},\chi)$ (for $r=0$, $(\id_1,1)$ is the identity anyon). We have the following fusion rule:
\begin{equation}
    (a_1,a_2,\cdots, a_{(N,r)},\id,\cdots,\id)\times (\id_g, \chi)=(\mb{a}_g,\chi).
\end{equation}
Here $\mb{a}=(a_1,a_2,\cdots, a_{(N,r)})$. Therefore $d_{(\mb{a}_{g^r},\chi)}=d_{\mb{a}}d_{(\id_{g^r},1)}$. 

To compute $d_{(\id_{g^r},1)}$, we use the fact that the total quantum dimension squared of all $g^r$ defects must be equal to $D_{\mathcal{B}}^2$~\cite{SET}. That is,
\[
\sum_{\mb{a}\in \mathcal{B}^{\boxtimes(N,r)}}d_\mb{a}^2 d_{\id_{g^r}}^2 = D_{\mathcal{B}}^{2(N,r)}d_{\id_{g^r}}^2= D_\mathcal{B}^{2N}.
\]
Thus $d_{\id_{g^r}}=D_{\mathcal{B}}^{N-(N,r)}$.

We also have the topological twist factors for the fluxes:
\begin{equation}
    \theta_{(\mb{a}_{g^r},\chi)}= \theta_{\mb{a}_{g^r}}\chi(g^r).
\end{equation}
Here $\theta_{\mb{a}_{g^r}}$ is the topological twist factor of the $\mb{a}_{g^r}$ defect. It can be chosen to be $\theta_{\mb{a}_{g^r}}=\theta_{\mb{a}}^{\frac{(N,r)}{N}}$, but we do not need its value in this section.

Now we consider the S matrix elements between fluxes and anyons. The entries of S matix could be computed from the topological spins and the fusion coefficients:
\begin{equation}
    S_{x,y}=\frac{1}{D}\sum_{z}N^{xy}_z d_z\frac{\theta_x \theta_y}{\theta_z}.
    \label{defS}
\end{equation}

We will use this to derive an important property of $S^{\mathcal{U}}_{(\mb{a}_{g^r},\chi_a),([\mb{b}]_1,\chi_b)}$, where $[\mb{b}]_1$ denotes a superposition anyon (i.e. $G_{\mb{b}}$ is smaller than $G$).  First we consider the fusion between a superposition anyon $([\mb{b}]_1,\chi_b)$ and a bare charge $(\id_1, \chi)$, where $\chi$ is a 1D rep. of $\Z_N$. Following Ref. [\onlinecite{SET}], we have
\begin{equation}\label{fusion:superposition}
    ([\mb{b}]_1,\chi_b) \times (\id_1, \chi)=([\mb{b}]_1,\chi_b\cdot \chi|_{G_{\mb{b}}}).
\end{equation}
Here $\chi|_{G_\mb{b}}$ is the restriction of $\chi$ to $G_\mb{b}$.

Suppose $\chi|_{G_{\mb{b}}}$ is trivial. For such a $\chi$, using \ref{fusion:superposition} we get:
\begin{equation}
    ([\mb{b}]_1,\chi_b) \times (\id_1, \chi)=([\mb{b}]_1,\chi_b),
\end{equation}
which implies that:
\begin{equation}
    (\mb{a}_{g^r},\chi_a)\times ([\mb{b}]_1,\chi_b) \times (\id_1, \chi)=(\mb{a}_{g^r},\chi_a)\times ([\mb{b}]_1,\chi_b).
    \label{fusionfixedpoint}
\end{equation}

Let us define ${\Lambda}_b$ as the group of $\chi$'s which restrict to identity on $G_{\mb{b}}$. It is easy to see that $\Lambda_b$ is all the irreps on $G/G_b$. Then Eq. \eqref{fusionfixedpoint} implies
\begin{equation}
    (\mb{a}_{g^r},\chi_a)\times ([\mb{b}]_1,\chi_b)=\sum_{\mb{c}_{g^r}} N_{\mb{a}_{g^r}, [\mb{b}]_1}^{\mb{c}_{g^r}}\sum_{\chi\in \Lambda_b}(\mb{c}_{g^r}, \chi_c\chi).
\end{equation}
While  $N_{\mb{a}_{g^r}, [\mb{b}]_1}^{\mb{c}_{g^r}}$ and $\chi_c$ are undetermined, they are not needed for our purpose.

Using the definition \eqref{defS}, we obtain
\begin{equation}
    \begin{split}    
    &S^{\mathcal{U}}_{(\mb{a}_{g^r},\chi_a),([\mb{b}]_1,\chi_b)}\\
    &=\frac{\theta_{(\mb{a}_{g^r},\chi_a)}\theta_{([\mb{b}]_1,\chi_b)}}{{D_{\mathcal{U}}}}
    \sum_{\mb{c}_{g^r}} N_{\mb{a}_{g^r}, [\mb{b}]_1}^{\mb{c}_{g^r}}d_{(\mb{c}_{g^r},\chi_c)}\sum_{\chi\in \Lambda_b} 
    \theta_{(\mb{c}_{g^r},\chi_c\chi)}^{-1}.
    \end{split}
\end{equation}
Let us consider the sum over $\chi\in \Lambda_b$, which can be simplified further to
\begin{equation}
    \begin{split}
     \sum_{\chi\in \Lambda_b} 
    \theta_{(\mb{c}_{g^r},\chi_c\chi)}^{-1}=  \theta_{(\mb{c}_{g^r},\chi_c)}^{-1} \sum_{\chi\in \Lambda_b} 
    \chi^{-1}(g^r).
    \end{split}
\end{equation}
First, we assume $g^r\in G_b$. By the definition of $\Lambda_b$, $\chi(g^r)=1$ so the sum evaluates to $|\Lambda_b|$. Then if $g^r\notin G_b$, the Schur's orthogonality theorem applied to $G/G_b$ shows that the sum should be 0. To summarize, we have shown that 
\begin{equation}
    S^{\mathcal{U}}_{(\mb{a}_{g^r},\chi_a),([\mb{b}]_1,\chi_b)}\propto \delta_{g^r\in G_\mb{b}}.
\end{equation}

Let us now consider what happens when $g^r\in G_{\mb{b}}$. It implies that $b$ has a period $(N,r)$, i.e. write $b=(b_1,b_2, \dots, b_N)$, then $b_i=b_{i+(N,r)\text{ mod }N}$. Thus for this caculation, we can group $(N,r)$ consecutive layers into one ``layer", with total number of layers $\tilde{N}=\frac{N}{(N,r)}$. It is also convenient to define $\tilde{\mb{b}}=(b_1,b_2,\cdots, b_{(N,r)})\in \tilde{\mathcal{B}}=\mathcal{B}^{\boxtimes (N,r)}$.  Therefore, the problem reduces to computing the $S$ matrix between a $(\mb{a}_{g^r},1)$ defect,  and $[\mb{b}]_1=(\tilde{\mb{b}},\dots,\tilde{\mb{b}})$. We will account for the charges later. 

Now we use a geometric picture to compute the S matrix. A cylinder with a $\mb{a}_{{g}^{{r}}}$ defect line can be viewed as a $\tilde{N}$-layer cylinder, with a $g^r$ cyclic permutation branch cut, which is topologically equivalent to a single-``layer" cylinder with the topological order described by $\tilde{\mathcal{B}}$. On the cylinder there is a Wilson line of the $\mb{a}$ anyon. We then compute the eigenvalue of the Wilson loop of $([\mb{b}]_1,1)$ along the other non-contractible cycle in two ways. Note that in the alternative picture, the Wilson loop of $([\mb{b}_1,1])$ becomes a Wilson loop of $\tilde{\mb{b}}$. We thus have:
\begin{equation}
    \frac{ S^{\mathcal{U}}_{(\mb{a}_{g^r},1),([\mb{b}]_1,1)}}{ S^{\mathcal{U}}_{(\mb{a}_{g^r},1),(\id_1,1)}} =\frac{S^{\tilde{\mathcal{B}}}_{\mb{a}\tilde{\mb{b}}}}{S^{\tilde{\mathcal{B}}}_{\mb{a} \id}}
\end{equation}
where $S^{\mathcal{U}}_{(\mb{a}_{g^r},1),(\id_1,1)}=\frac{d_{(\mb{a}_{g^r},1)}}{D_{\mathcal{U}}}$, $S^{\mathcal{B}}_{a \id}=\frac{d_\mb{a}}{D_{\tilde{\mathcal{B}}}}=\frac{d_\mb{a}}{D_{\mathcal{B}}^{(N,r)}}$. Using $d_{(\mb{a}_{g^r},1)}=D_{\mathcal{B}}^{N-(N,r)}d_\mb{a}$ and $D_{\mathcal{U}}=ND_{\mathcal{B}}^N$, we have:
\begin{equation}
\begin{split}
    S^{\mathcal{U}}_{(\mb{a}_{g^r},1),([\mb{b}]_1,1)}&= \frac{d_\mb{a}D_B^{N-(N,r)}}{ND_{\mathcal{B}}^N}\frac{D_{\mathcal{B}}^{(N,r)}}{d_\mb{a}}S^{\tilde{\mathcal{B}}}_{\mb{a}\tilde{\mb{b}}}\\
    &= \frac{1}{N}S^{\tilde{\mathcal{B}}}_{\mb{a}\tilde{\mb{b}}}.
\end{split}
\end{equation}

Including the contribution from the charge $\chi_b$, we find
\begin{equation}
     S^{\mathcal{U}}_{(\mb{a}_{g^r},\chi_a),([\mb{b}]_1,\chi_b)}= \frac{1}{N}\chi_b(g^r) S^{\tilde{\mathcal{B}}}_{\mb{a}\tilde{\mb{b}}}\delta_{g^r\in G_{\mb{b}}}.
\end{equation}

Now we study gapped boundaries of $\mathcal{B}^{\boxtimes N}$ from the gauging perspective. Suppose we have a Lagrangian algebra $\cal{A}_0$ of $\cal{B}^{\boxtimes N}$. It can be ``lifted" to a condensable algebra  $\cal{A}$ in the gauged theory $\cal{U}$. Namely, we can first break $\cal{A}_0$ into orbits under $G$, and each orbit becomes a single anyon after gauging. 
\begin{equation}
    \cal{A}_0=\sum_{[\mb{a}]}n_{\mb{a}}\sum_{\mb{b}\in [\mb{a}]}\mb{b}.
\end{equation}
Here $[\mb{a}]$ is an orbit under $G$, with $\mb{a}$ being a representative element. After gauging, it becomes
\begin{equation}
    \cal{A}=\sum_{[\mb{a}]}n_{\mb{a}}([\mb{a}]_{1},\chi_\mb{a}).
\end{equation}
Roughly speaking, $\chi_{\mb{a}}$ is the symmetry charge carried by the $\mb{a}$ anyon when it condenses.
The assignment of $\chi_\mb{a}$ is not arbitrary.  In fact, because $U$ and $\eta$ are all $1$ in our example, the $\chi_\mb{a}$'s are precisely those defined in Sec. \ref{sec:charged-condensate}.

 By definition, condensing $\cal{A}_0$ in $\cal{B}^{\boxtimes N}$ results in a trivial theory (the Vec MTC). If the condensation preserves the $\Z_N$ symmetry, condensing $\cal{A}$ in $\cal{U}$ should result in a deconfined $\Z_N$ gauge theory, denoted by $\cal{D}$. In the same convention, anyons in $\cal{D}$ will be denoted by $(g^r, \chi)$. The vacuum is $\id\equiv (1, 1)$. Note that depending on the choice of the charges $\chi_{\mb{a}}$'s, the $\Z_N$ gauge theory may be twisted. We are in particular interested in whether there exists a choice of $\chi$'s such that the $\Z_N$ gauge theory is not twisted.

 We briefly recall a few basic facts about anyon condensation from a MTC $\cal{U}$ to $\cal{D}$. It is useful to think of the anyon condensation as defining a gapped interface between the theories $\cal{U}$ and $\cal{D}$. To fully describe this process, it is also necessary to define the theory of (possibly confined) excitations on the interface, called $\cal{T}$. $\cal{D}$ is the ``deconfined" subcategory of $\cal{T}$. 

 The relations between the three theories are encoded in the restriction map $r$ and the lifting map $l$.
  An anyon $\alpha$ in the $\cal{U}$ theory can be ``restricted" to the $\cal{T}$ theory:
\begin{equation}
    r(\alpha)=\sum_{t\in \cal{T}}n_{\alpha,t}t.
\end{equation}
where the $n_{\alpha,t}$'s are non-negative integers.
On the other hand, a particle in the $\cal{T}$ can be lifted back to the $\cal{U}$ theory:
\begin{equation}
    l(t)=\sum_{\alpha \in \cal{U}}n_{\alpha,t}\alpha.
\end{equation}
Clearly $l(\id)$ is the Lagrangian algebra $\cal{A}$. We are mostly interested in $n_{\alpha,t}$'s when $t\in \cal{D}$.

 The integers $n_{\alpha,t}$ should satisfy various constraints. We will only need the following condition: the restriction/lifting maps should commute with the modular matrices of $\cal{U}$ and $\cal{D}$. More specifically, 
 for $\alpha\in \cal{U}$ and $t\in \cal{D}$, we have
 \begin{equation}
     \sum_{\beta\in \cal{U}}S_{\alpha\beta}n_{\beta,t}=\sum_{s\in \cal{D}}n_{\alpha,s}S_{st}.
     \label{Sn=nS}
 \end{equation}
Secondly, for a given $t\in \cal{U}$,  all $\alpha$'s with $n_{\alpha,t}\neq 0$ share the same topological twist factor and $\theta_{\alpha}=\theta_t$.
 
  We will denote 
 \begin{equation}
     l((g,1))=\sum_{a\in \mathcal{B}} w_{g}(a)(a_{g},1),
 \end{equation}
 where $w_{g}(a)$ is a non-negative integer. We shall use \eqref{Sn=nS} to constrain $w_g(a)$, which can then determine the topological twist factor of $(g,1)$.

Setting $\alpha=(a_g, 1)$, $t=\id$ in \eqref{Sn=nS}, we have the left-hand side given by 
\begin{equation}
    \sum_{([\mb{b}]_1,\chi_b)\in \mathcal{A}} S^{\cal{U}}_{(a_g,1),([\mb{b}]_1,\chi_\mb{b})}n_{\mb{b}} = \frac{1}{N}\sum_{\mb{b}=g(\mb{b})}\chi_\mb{b}S_{ab}^{\cal{B}}n_{\mb{b}}.
\end{equation}
The RHS is $\frac{1}{N}w_g(a)$. So we find
\begin{equation}
    \sum_{\mb{b}=g(\mb{b})}\chi_\mb{b}S^{\cal{B}}_{ab}n_{\mb{b}} = w_g(a).
\end{equation}
Here $\mb{b}=g(\mb{b})$ means $\mb{b}=(b,b,\cdots,b)$.
Similar relations can be derived for other $g^r$. Given that the S matrix elements are generally not integers, it is a nontrivial consistency check that the sum on the left-hand side yields a non-negative integer. 

Let us now apply the relation to $\cal{B}=\text{Spin}(2n)_1$ theory. In this case, $b$ runs over all anyons in $\mathcal{B}$ and with $n_\mb{b}=1$ we have
\begin{equation}
    \sum_{b\in \cal{B}}S_{ab}^{\cal{B}}\chi_{\mb{b}}=w_g(a).
\end{equation}
We have shown in that $\chi_{\mb{b}}=1$ for all $\mb{b}=g(\mb{b})$ is allowed for $n=1,2$ and $4$, and in fact for all other values of $n$. With this choice, the unitarity of S matrix implies that $w_g(a)=2\delta_{a,\id}$. Therefore, $\theta_{(g,1)}=\theta_{(\id_g,1)}=1$ and the $\Z_N$ gauge theory is not twisted.

In the case of the Ising theory with $N=16$, the LHS becomes
\begin{equation}
\begin{split}
    \frac12(1+128\sqrt{2}\chi_\sigma + \chi_\psi)=w_g(\id),\\
    \frac12(1-128\sqrt{2}\chi_\sigma+\chi_\psi)=w_g(\psi),\\
    \frac{1}{\sqrt{2}}(1-\chi_\psi)=w_g(\sigma).
\end{split}
\end{equation}
Regardless of the choice of $\chi_\sigma$ and $\chi_\psi$, $w_g(\id)$ and $w_g(\psi)$ can not be an integer, and the only possible integer value for $w_g(\sigma)$ is 0. We conclude that such a symmetry-preserving condensation in Ising$^{\boxtimes 16}$ does not exist. This confirms the more heuristic argument in Sec. \ref{Ising}.

\bibliography{4dspt}
\end{document}